\documentclass[pdflatex,sn-apa,iicol]{sn-jnl}

\usepackage{mathrsfs}%
\usepackage{graphicx}%
\usepackage{multirow}%
\usepackage{amsmath,amssymb,amsfonts}%
\usepackage{amsthm}%

\usepackage[title]{appendix}%
\usepackage{xcolor}%
\usepackage{textcomp}%
\usepackage{manyfoot}%
\usepackage{booktabs}%
\usepackage{algorithm}%
\usepackage{algorithmicx}%
\usepackage{algpseudocode}%
\usepackage{listings}%

\usepackage{enumitem}
\usepackage{array}
\usepackage[table]{xcolor}
\usepackage{rotating}
\usepackage{threeparttable}
\usepackage{tabularx}
\usepackage{pifont}
\usepackage{makecell}
\usepackage{placeins}

\theoremstyle{thmstyleone}%
\theoremstyle{thmstyletwo}%

\theoremstyle{thmstylethree}%

\begin{document}

\title[Article Title]{End-to-End Learned Image Coding: From Rate--Distortion Optimization to Practical Deployment}


% \author*[1,2]{\fnm{First} \sur{Author}}\email{iauthor@gmail.com}



% \affil*[1]{\orgdiv{Department}, \orgname{Organization}, \orgaddress{\street{Street}, \city{City}, \postcode{100190}, \state{State}, \country{Country}}}

% \affil[2]{\orgdiv{Department}, \orgname{Organization}, \orgaddress{\street{Street}, \city{City}, \postcode{10587}, \state{State}, \country{Country}}}

% \affil[3]{\orgdiv{Department}, \orgname{Organization}, \orgaddress{\street{Street}, \city{City}, \postcode{610101}, \state{State}, \country{Country}}}

\author{\fnm{Ming} \sur{Lu}}%\email{minglu@nju.edu.cn}
\equalcont{These authors contributed equally to this work.}

\author{\fnm{Junqi} \sur{Shi}}%\email{junqishi@smail.nju.edu.cn}
\equalcont{These authors contributed equally to this work.}

\author*{\fnm{Zhan} \sur{Ma}*}\email{mazhan@nju.edu.cn}

\affil{\orgdiv{School of Electronic Science and Engineering}, \orgname{Nanjing University}, \orgaddress{\city{Nanjing}, \postcode{210023}, \state{Jiangsu}, \country{China}}}


\abstract{{End-to-end learned image coding (LIC) has achieved substantial improvements in rate--distortion (R--D) performance, yet practical deployment requires addressing a broader set of system-level constraints. This article provides a deployment-oriented review of LIC, with particular attention to the transition from high-performance neural compression models to practical codecs. We first summarize the basic end-to-end coding framework, including learned transforms, quantization, entropy modeling, and R--D optimization. We then review practical extensions that are increasingly important for deployment, including variable-rate coding and rate control, integer-precision inference and cross-platform consistency, robustness and communication resilience, lightweight and hardware-aware implementation, and standardization and interoperability. To study these issues under a common setting, we further introduce TinyLIC as an open benchmark and reproducible case study for rate--distortion--complexity analysis. Experimental results illustrate the trade-offs among compression efficiency, computational complexity, numerical consistency, and robustness. In particular, TinyLIC with approximately 10 KMACs/pixel decoding complexity substantially outperforms the H.265/HEVC image-profile anchor on Kodak, while the 50 KMACs/pixel configuration achieves performance comparable to the H.266/VVC image-profile anchor on several test sets. The review highlights practical constraints and design trade-offs that remain central to the deployment and interoperability of learned image codecs. Materials are accessible at \url{https://njuvision.github.io/TinyLIC}.}}




\maketitle

\section{Introduction}\label{sec:introduction}

Image coding has long been a cornerstone of visual communication systems~\citep{shannon1948mathematic, ivins1969prints}. It supports a wide range of visual communication and storage applications, including social photo sharing, digital archiving, mobile imaging, and remote sensing. For more than three decades, mainstream image coding standards have largely followed a modular and rule-driven design philosophy~\citep{wallace1991jpeg, bellard2016bpg, vvc_intra}. Representative codecs improve compression efficiency by refining individual components such as transform~\citep{clarke1985transform}, prediction~\citep{kobayashi1974image}, quantization~\citep{gersho1978principles}, and entropy coding~\citep{senoo1990vector}, while maintaining interoperability through carefully specified bitstream syntax and standardization procedures. This paradigm has achieved remarkable success, but its components are mostly hand-crafted and optimized within a constrained design space.

Over the past decade, image compression has undergone a paradigm shift driven by deep learning. By learning compact latent representations directly from data, end-to-end learned image coding (LIC) frameworks~\citep{chen2021end, he2022elic, lu2022transformer, liu2023learned, duan2023qarv} have surpassed the Joint Photographic Experts Group (JPEG) standard~\citep{wallace1991jpeg}, JPEG 2000~\citep{JPEG2K}, and even image-profile anchors derived from modern video coding standards, including Better Portable Graphics (BPG, H.265/High Efficiency Video Coding (HEVC) image-profile anchor) and VTM Intra (H.266/Versatile Video Coding (VVC) image-profile anchor)~\citep{bellard2016bpg, vvc_intra}, on many public benchmarks in terms of rate--distortion (R--D) performance~\citep{hu2021learning, ding2021advances}. At the same time, the flexibility of learned representations has extended image coding beyond conventional pixel-fidelity optimization toward perceptual reconstruction, ultra-low-bitrate communication~\citep{keultra}, and machine-oriented visual representation~\citep{duan2023unified}.

However, as LIC matures, the field's primary bottleneck is shifting. The central question is no longer only whether learned codecs can achieve better R--D curves, but also whether they can operate as reliable, controllable, efficient, and interoperable compression subsystems in real communication environments. Practical deployment imposes constraints that are often underrepresented in academic benchmarks, including computational complexity, decoding latency, rate-control flexibility, robustness to content variation and perturbations, and compatibility with heterogeneous hardware platforms~\citep{sze2017hardware}.

This shift has redirected attention from performance-driven model design to system-oriented codec design. Recent studies have explored lightweight architectures~\citep{bao2025shiftlic}, low-complexity entropy models~\citep{he2021checkerboard}, variable-rate and multi-rate coding~\citep{chen2020variable}, robustness under distribution shifts and adversarial perturbations~\citep{chen2023toward}, model quantization and integer inference~\citep{sun2022q}, and hardware-friendly implementations~\citep{chen2024exploring}. In parallel, the standardization of JPEG Artificial Intelligence (JPEG AI)\footnote{\url{https://jpeg.org/jpegai/}}~\citep{JPEG-AI} and IEEE 1857.11\footnote{\url{https://standards.ieee.org/ieee/1857.11/10691/}}~\citep{IEEE1857} indicates that LIC is moving from academic prototypes toward industrial validation. These developments motivate a survey that treats compression efficiency, deployability, and interoperability as coupled design objectives.

Existing surveys have provided valuable perspectives on this rapidly evolving field. Early reviews focused on the transition from traditional coding tools to neural-network-based or end-to-end compression~\citep{ma2019image,liu2020deep,ding2021advances}, while later studies offered increasingly comprehensive coverage of learned transforms, entropy models, perceptual coding, rate control, and benchmark performance~\citep{hu2021learning,mishra2022deep,jamil2023learning,zhang2023survey,chen2024generative,chen2024survey,huang2024unveiling}. These reviews provide valuable algorithmic and conceptual context, but deployment-oriented issues such as integer inference, decoder consistency, entropy-coder runtime, robustness, rate-control guarantees, lightweight implementation, and standard conformance are usually discussed separately or only as future challenges.

This article therefore reviews LIC from a deployment-oriented perspective. Rather than providing another exhaustive taxonomy of neural architectures, we briefly summarize the core transform–quantization–entropy–optimization framework and focus primarily on the capabilities required to convert learned compression models into practical codecs. In particular, we consider five closely related aspects: \emph{rate flexibility and control}, which determine how a codec adapts to varying bitrate requirements; \emph{model quantization and cross-platform consistency}, which affect low-precision implementation and deterministic decoding; \emph{robustness and communication resilience}, which characterize codec behavior under perturbations and transmission errors; \emph{lightweight and hardware-aware deployment}, which determines computational feasibility on resource-constrained platforms; and \emph{standardization and interoperability}, which provide the basis for reproducible and compatible codec implementations.

To complement the literature review, we further introduce {\emph TinyLIC} as an open benchmark and reproducible case study for practical LIC under a shared architecture. TinyLIC provides a lightweight and configurable reference framework in which computational complexity and representative deployment functions can be systematically varied. The framework supports multiple complexity levels together with variable-rate coding, quantization-aware consistent decoding, adversarially robust training, progressive transmission, and network-aware loss-resilient coding. This design enables controlled analysis of how additional system requirements affect rate, distortion, complexity, and robustness, and provides a common setting for observing trade-offs that are difficult to isolate when comparing independently developed codecs.

The main contributions of this survey are summarized as follows.
{\begin{itemize}
    \item {\bf A deployment-oriented view of LIC.} We organize the field from the perspective of the transition from R--D-optimized neural compression models to practical codecs, while treating the core LIC architecture as a concise algorithmic foundation rather than the primary focus.
    \item {\bf A systematic review of practical codec capabilities (Sec.~\ref{sec:practical}).} We examine rate flexibility and control, model quantization and cross-platform consistency, robustness and transmission resilience, lightweight and hardware-aware implementation, and standardization, with particular attention to the interactions and trade-offs among these requirements.
    \item {\bf A reproducible benchmark for practical LIC (Sec.~\ref{sec:tinylic}).} We introduce TinyLIC as a lightweight and extensible reference framework for rate--distortion--complexity analysis and for controlled evaluation of variable-rate coding, consistent decoding, progressive transmission, adversarial robustness, and loss resilience.
    \item {\bf A discussion of emerging directions beyond conventional R--D optimization (Sec.~\ref{sec:conclusion}).} We identify open challenges in hardware-efficient and interoperable learned codecs and discuss longer-term developments toward coding for machine intelligence, neural scene representations, and foundation-model-assisted compression.
\end{itemize}}

\begin{figure}[t]
\centering
\includegraphics[width=\linewidth]{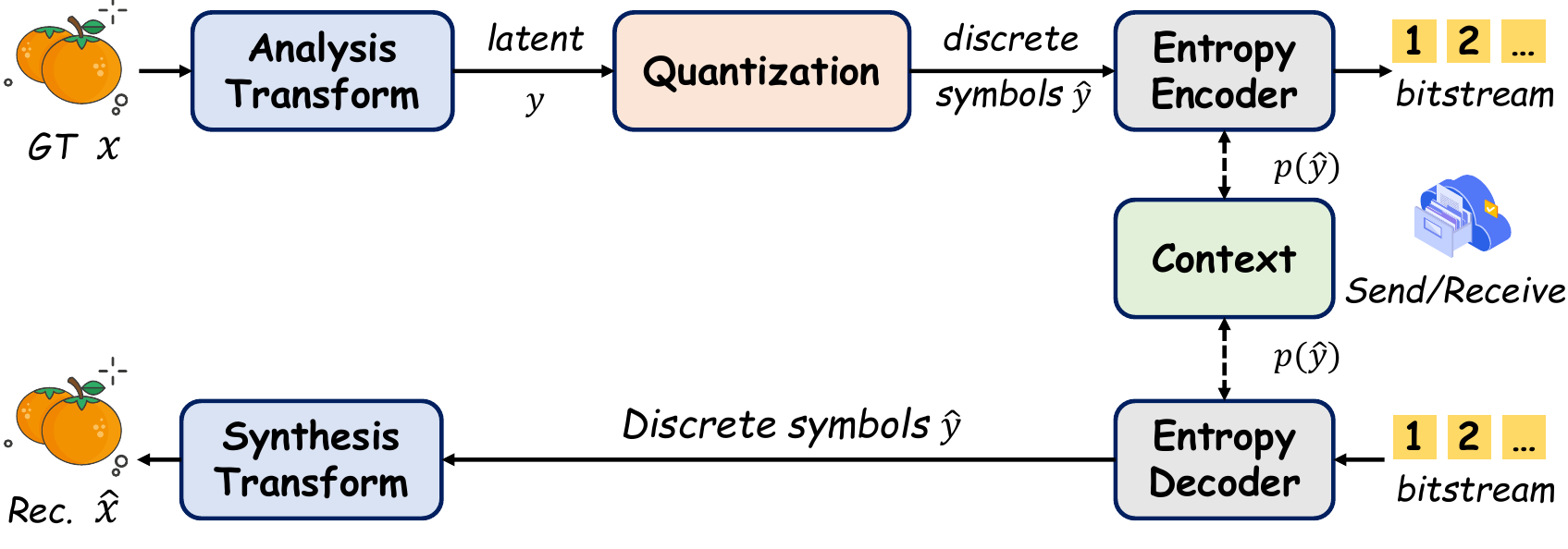}
\caption{A variational-autoencoder (VAE)-based LIC. The analysis transform maps the input image into latents that are then quantized and entropy-coded. The decoder reconstructs the image using the entropy-decoded latents and the synthesis transform.}
\label{fig:arch}
\end{figure}

\section{Core Framework: A Brief Overview}
\label{sec:core_framework}

\begin{figure*}[t]
\centering
\includegraphics[width=0.96\linewidth]{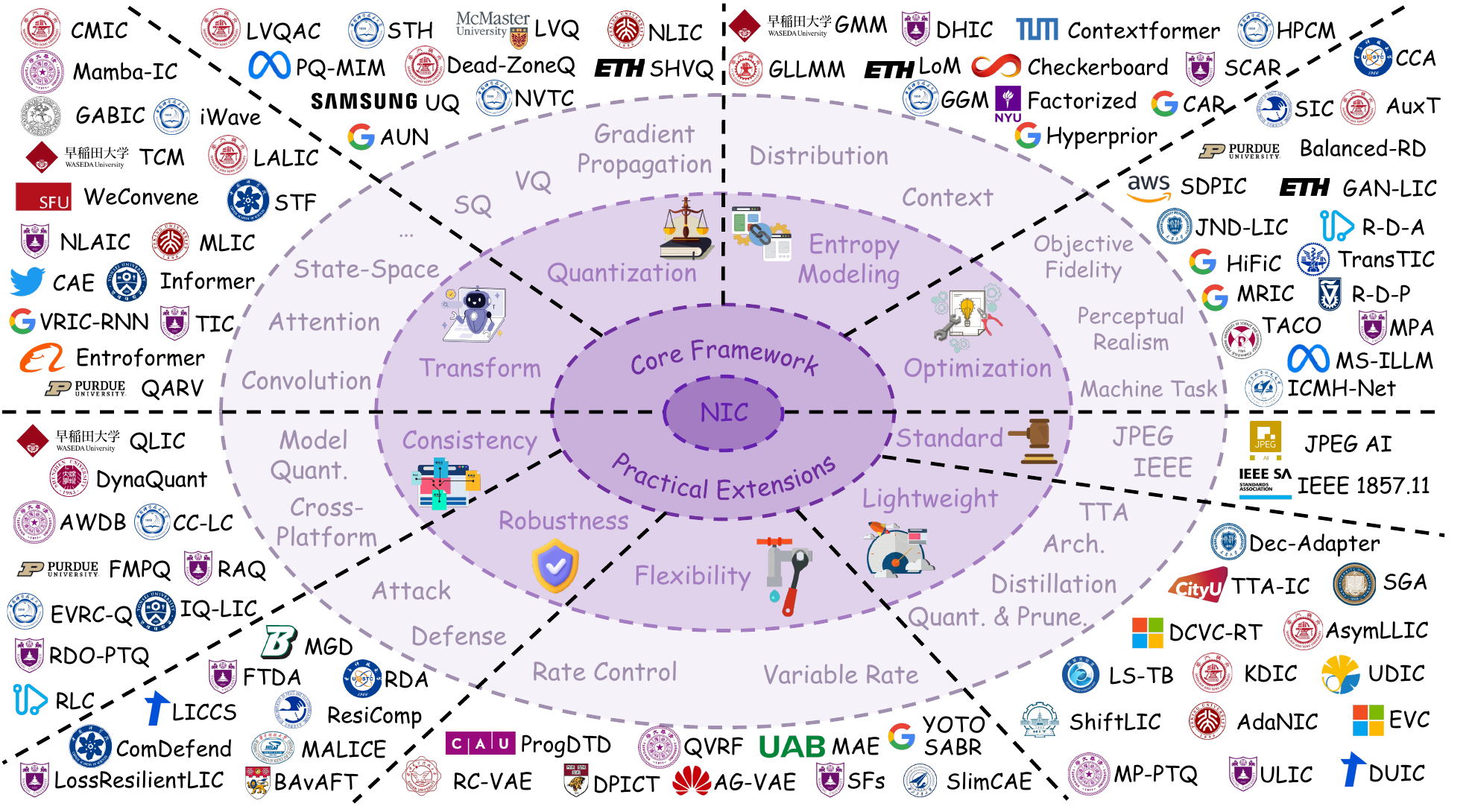}
\caption{Taxonomy of end-to-end learned image coding from algorithmic foundations to practical deployment. The core framework consists of learned transforms, quantization, entropy modeling, and optimization objectives, while practical extensions address rate flexibility, numerical consistency, robustness, lightweight deployment, and standardization. The latter form the primary focus of this survey.}
\label{fig:taxonomy}
\end{figure*}

End-to-end learned image coding (LIC) follows the classical principle of transform coding while replacing hand-crafted coding modules with jointly optimized neural components. As shown in Fig.~\ref{fig:arch}, given an input image $x$, an analysis transform $f_{\theta}$ maps the image into a compact latent representation 
\begin{align}
    y=f_{\theta}(x),
\end{align}
which is discretized into $\hat{y}=Q(y)$ and compressed according to a learned probability model $p_{\psi}(\hat{y})$. At the decoder, a synthesis transform $g_{\phi}$ reconstructs the image as
\begin{align}
    \hat{x}=g_{\phi}(\hat{y}).
\end{align}

Most modern LIC methods instantiate this formulation within a
variational-autoencoder (VAE) framework \citep{kingma2013auto,balle2017end,balle2018variational}. Auxiliary latent variables, such as hyperpriors, are commonly introduced to convey side information for probability estimation \citep{balle2018variational}. The expected coding rate is approximated by the negative log-likelihood of the transmitted symbols, while reconstruction quality is measured by a distortion function $D({x},\hat{{x}})$. The codec is typically optimized through the Lagrangian rate--distortion (R--D) objective
\begin{equation}
    \mathcal{L}
    =
    R
    +
    \lambda D({x},\hat{{x}}),
\end{equation}
where $R$ denotes the expected bitrate and $\lambda$ controls the trade-off between compression rate and reconstruction quality. This formulation provides a common foundation on which different transforms, quantizers, entropy models, and reconstruction objectives can be developed \citep{balle2020nonlinear}.

As summarized in Fig.~\ref{fig:taxonomy}, the development of LIC can be viewed from two complementary perspectives. The \emph{core framework} consists of four tightly coupled components: transform, quantization, entropy modeling, and optimization. The \emph{practical extensions}, reviewed in Sec.~\ref{sec:practical}, concern the system properties required beyond the basic R--D formulation, including rate flexibility, numerical consistency, robustness, lightweight deployment, and standardization.

\begin{table*}[t]
  \centering
  \tiny
  \caption{Overview of representative learned image compression methods.}
  \label{tab:lic_overview}
  \setlength{\tabcolsep}{2.0pt}
  \renewcommand{\arraystretch}{1.10}

  \begin{tabularx}{\textwidth}{
    >{\raggedright\arraybackslash}X
    >{\centering\arraybackslash}p{2.0cm}
    >{\centering\arraybackslash}p{1.5cm}
    >{\centering\arraybackslash}p{2.0cm}
    >{\centering\arraybackslash}p{0.75cm}
    >{\centering\arraybackslash}p{1.20cm}
    >{\centering\arraybackslash}p{1.25cm}
    >{\centering\arraybackslash}p{1.25cm}
  }
    \toprule
    \rowcolor{gray!15}
    \textbf{Method} &
    \textbf{Pub.} &
    \multicolumn{3}{c}{\textbf{Core Components}} &
    \makecell{\textbf{Params}\\\textbf{(M) $\downarrow$}} &
    \makecell{\textbf{KMACs/}\\\textbf{pixel $\downarrow$}} &
    \makecell{\textbf{BD-rate}\\\textbf{(\%) $\downarrow$}} \\
    \rowcolor{gray!15}
    &
    &
    \textbf{Arch.} &
    \textbf{Context} &
    \textbf{Dist.} &
    &
    &
    \\
    \midrule

    Factorized~\citep{balle2017end}
      & ICLR'17
      & CNN & Factorized & UM
      & 7.03 & 204.00 & 67.80 \\

    Hyperprior~\citep{balle2018variational}
      & ICLR'18
      & CNN & Hyperprior & GM
      & 11.81 & 208.97 & 30.14 \\

    JointAR~\citep{minnen2018joint}
      & NeurIPS'18
      & CNN & SAR & GM
      & 25.50 & 224.80 & 10.61 \\

    ChannelAR~\citep{minnen2020channel}
      & ICIP'20
      & CNN & CAR & GM
      & 28.72 & 243.00 & 7.22 \\

    NLAIC~\citep{chen2021end}
      & TIP'20
      & Attention & SCAR (3D) & GM
      & 65.50 & 823.54 & 5.92 \\

    GMM~\citep{cheng2020learned}
      & CVPR'20
      & CNN & SAR & GMM
      & 29.63 & 512.80 & 4.55 \\

    ELIC~\citep{he2022elic}
      & CVPR'22
      & CNN & SCAR & GM
      & 36.93 & 573.88 & -3.22 \\

    TIC~\citep{lu2022transformer}
      & DCC'22
      & Attention & SCAR & GM
      & 28.40 & 506.72 & -4.58 \\

    TCM~\citep{liu2023learned}
      & CVPR'23
      & Attention & CAR & GM
      & 76.57 & 1823.58 & -10.70 \\

    QARV~\citep{duan2023qarv}
      & TPAMI'24
      & CNN & HAR & GM
      & 93.40 & 718.96 & -5.81 \\

    FLIC~\citep{li2024frequency}
      & ICLR'24
      & Attention & CAR & GM
      & 70.97 & 1096.04 & -12.97 \\
    
    MLIC++~\citep{jiang2025mlic++}
      & TOMM'25
      & CNN & SCAR & GM
      & 116.72 & 1282.81 & -11.83 \\
      
    MambaIC~\citep{zeng2025mambaic}
      & CVPR'25
      & Mamba & SCAR & GM
      & 75.78 & 1284.86 & -15.72 \\

    HPCM~\citep{li2025learned}
      & ICCV'25
      & CNN & HAR & GM
      & 89.71 & 1261.29 & -19.19 \\

    DHIC~\citep{cong2026taming}
      & ICLR'26
      & CNN & HAR & GM
      & 106.93 & 977.73 & {-19.73} \\

    \bottomrule
  \end{tabularx}

  \vspace{2pt}
  \begin{minipage}{\textwidth}
  \scriptsize
  \textit{Note:}
  \textbf{Pub.}: publication venue and year;
  \textbf{Arch.}: main architecture;
  \textbf{Dist.}: distribution model.
  Because BD-rate and complexity values depend on anchor version, color pipeline, test-set split, interpolation protocol, and whether results are reproduced or reported by the original papers, the table should be read as a normalized high-level comparison rather than a bit-exact cross-paper leaderboard.
  \end{minipage}
\end{table*}

\textbf{Transform.} 
Learned transforms determine how visual information is mapped into compact latent representations. Early end-to-end codecs were dominated by convolutional architectures \citep{balle2017end,balle2018variational}, which provide efficient local feature extraction and remain attractive for practical implementations. Subsequent methods introduced non-local attention and Transformer-based architectures \citep{chen2021end,lu2022transformer,liu2023learned,li2024frequency} to capture longer-range spatial dependencies, while more recent studies have explored state-space models \citep{zeng2025mambaic} and hierarchical latent representations \citep{duan2023qarv,li2025learned,cong2026taming}. These developments improve representation capacity, but also lead to different computational, memory, and parallelization characteristics. Transform design therefore increasingly reflects a trade-off between representation capability and implementation efficiency rather than compression performance alone.

\textbf{Quantization.}
Quantization connects continuous neural representations with discrete symbols that can be entropy coded. Scalar quantization remains the dominant choice because of its simplicity and compatibility with practical entropy coders. Since hard discretization is non-differentiable, early learned codecs typically rely on differentiable approximations such as additive uniform noise during training \citep{theis2017lossy,balle2017end}. Alternative strategies include universal quantization and soft-to-hard relaxation \citep{agustsson2017soft,agustsson2020universally,guo2021soft}. More expressive vector or structured quantization schemes can exploit dependencies among latent dimensions, but generally introduce additional optimization or search complexity. It is important to distinguish the signal quantization considered here from \emph{model quantization}, which converts network weights and activations to low-precision representations for deployment and is discussed separately in Sec.~\ref{sec:quantization_consistency}.

\textbf{Entropy Modeling.}
Compression efficiency further depends on how accurately the probability distribution of quantized latents is modeled. Early end-to-end methods used factorized latent distributions \citep{balle2017end}, followed by hyperprior-based models \citep{balle2018variational} that transmit auxiliary side information to capture spatially varying latent statistics. Autoregressive models subsequently exploit dependencies among previously decoded symbols and can operate along spatial, channel, or joint spatial-channel dimensions \citep{minnen2018joint,minnen2020channel,chen2021end}. More recent approaches combine richer context models with hierarchical representations to further improve probability estimation \citep{jiang2025mlic++,li2025learned,cong2026taming}. Stronger entropy models generally reduce statistical redundancy, but often introduce additional decoding steps, memory accesses, or sequential dependencies. The balance between probability-modeling accuracy and decoding efficiency is therefore particularly important for practical deployment.

\textbf{Optimization Objective.}
The fourth component is the optimization objective. Classical learned codecs primarily optimize signal fidelity using distortion measures such as mean squared error (MSE), resulting in the conventional R--D formulation \citep{balle2017end,balle2020nonlinear}. The same end-to-end framework can, however, accommodate other notions of reconstruction utility. Perceptual compression introduces feature-space or distribution-level objectives to improve visual realism at low bitrates, reflecting the rate--distortion--perception trade-off \citep{blau2018perception,blau2019rethinking,mentzer2020high}. Machine-oriented coding further incorporates task-related objectives so that compressed representations retain information useful for recognition or analysis \citep{chamain2021end,zhang2024all}. These extensions broaden the objective of learned coding from faithful signal reconstruction toward perceptually and semantically useful representation, although their evaluation criteria and deployment requirements remain less standardized than those of fidelity-oriented coding.

The evolution of these components has substantially improved compression efficiency. Table~\ref{tab:lic_overview} provides a representative snapshot of this progression. Factorized and hyperprior models established the early VAE-based framework, while autoregressive context models, attention-based transforms, hierarchical representations, and more expressive probability models progressively improved R--D performance. At the same time, the table reveals an important characteristic of modern LIC: improved coding performance is frequently accompanied by increased model size and computational complexity. Recent high-performance codecs commonly require several hundred to more than one thousand KMACs/pixel, whereas practical systems may operate under substantially tighter computation and memory budgets. Because complexity measurements are also affected by implementation
details, entropy-coding procedures, hardware platforms, and evaluation protocols, these values should be interpreted primarily as an architecture-level trend rather than a bit-exact cross-paper leaderboard.

This observation marks an important transition in LIC. The transform-quantization-entropy-optimization framework has become increasingly mature, and further improvements in R--D performance alone do not guarantee practical usability. A deployable codec must additionally determine how bitrate is controlled, whether decoding remains numerically consistent across heterogeneous platforms, how the codec behaves under input or transmission perturbations, whether its computation maps efficiently to available hardware, and how interoperability is maintained across implementations.

These properties are closely coupled with the underlying coding framework. For example, stronger autoregressive dependencies may improve entropy modeling but increase decoding latency and error propagation; low-precision inference may improve hardware efficiency while altering R--D behavior; and robustness-oriented mechanisms may sacrifice part of the coding efficiency to improve reliability. Practical LIC is therefore better viewed as a multi-objective system-design problem rather than an extension of
R--D optimization alone.

{For this reason, the remainder of this survey treats the core LIC framework as the algorithmic foundation rather than the primary organizing dimension. Section~\ref{sec:practical} focuses on the practical extensions required to turn an R--D-optimized neural compression model into a deployable codec, including rate flexibility and control, model quantization and cross-platform consistency, robustness and communication resilience, lightweight and hardware-aware deployment, and standardization and interoperability.}

\begin{figure*}[t]
\centering
\includegraphics[width=0.95\linewidth]{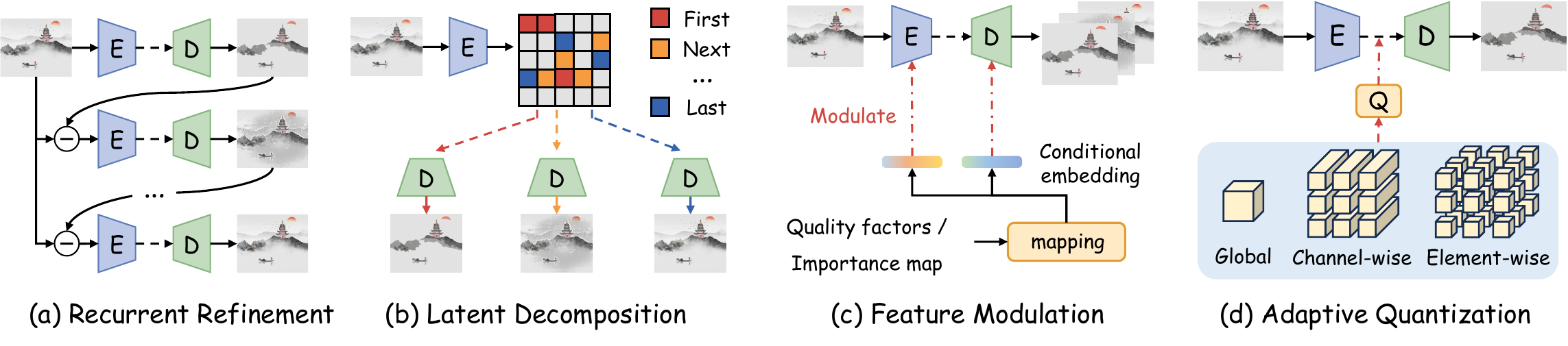}
\caption{Representative mechanisms for variable-rate learned image coding. Rate adaptability can be introduced through recurrent refinement, latent decomposition, feature modulation, or adaptive quantization. These mechanisms differ in achievable rate granularity, inference complexity, and compatibility with a fixed codec topology.}
\label{fig:variable_rate}
\end{figure*}

\section{Practical Extensions for Deployable LIC}\label{sec:practical}

% Toward real-world deployment, both academia and industry have devoted substantial efforts to improving flexibility, controllability, and system efficiency. In this section, we review practical extensions that enable learned codecs under operational constraints encountered in communication, storage, and edge-intelligence pipelines.

The core framework reviewed in Sec.~\ref{sec:core_framework} defines how learned codecs represent, quantize, and compress visual information. Practical deployment, however, imposes requirements that are not fully characterized by nominal R--D performance. A deployable codec must adapt to varying bitrate constraints, produce deterministic results across heterogeneous platforms, remain reliable under content and transmission perturbations, operate within computation and memory budgets, and maintain interoperability across implementations. These requirements arise at different levels of the coding pipeline and are often coupled with the underlying transform, entropy model, and optimization objective.

Accordingly, we organize practical LIC around five closely related dimensions. \emph{Rate flexibility and control} determine whether a codec can adapt its operating point to bandwidth or quality requirements. \emph{Model quantization and cross-platform consistency} address low-precision execution and deterministic entropy decoding. \emph{Robustness and communication resilience} characterize codec behavior under input perturbations, distribution shifts, and transmission errors. \emph{Lightweight and hardware-aware deployment} concerns computational cost, memory, latency, and hardware efficiency. Finally, \emph{standardization and interoperability} provide common bitstream, decoder, and conformance requirements for practical adoption. Rather than treating these capabilities as independent add-on modules, this section emphasizes their interactions with compression efficiency and with one another.

\subsection{Rate Flexibility and Control}
\label{sec:rate_flex}

% Variable-rate coding (VRC) and rate control are essential for practical image codecs, as they enable a single model to adapt to varying bandwidth constraints and user preferences without retraining or maintaining multiple models.

Practical LIC must adapt to varying bandwidth, storage, and quality
requirements without maintaining multiple independently trained models.
Three related capabilities are particularly important. \emph{Variable-rate
coding} allows a single codec to cover multiple R--D operating points;
\emph{progressive coding} further requires these operating points to be
nested within a common bitstream; and \emph{rate control} selects coding
parameters so that the actual bitrate approaches a prescribed target.
We review these three capabilities in turn.

\subsubsection{Variable-rate Coding}
\label{sec:variable_rate}

% Variable-rate coding aims to support multiple R--D operating points within a unified framework. Unlike traditional codecs, where bitrate is primarily controlled through explicit quantization parameter (QP) adjustment, learned image compression requires dedicated mechanisms to regulate bitrate while maintaining stable reconstruction quality. Existing approaches mainly differ in how and where rate adaptability is introduced into the codec pipeline.

Variable-rate coding (VRC) aims to span a range of R--D operating points within a single model, rather than training an independent network for each bitrate or quality level. In contrast to conventional codecs, where bitrate can often be adjusted directly through a quantization parameter (QP), learned codecs couple bitrate with latent representations, entropy statistics, and network activations. Rate adaptation therefore requires an explicit mechanism for modifying the coding behavior of the learned model. As illustrated in Fig.~\ref{fig:variable_rate}, existing approaches can be broadly grouped into recurrent refinement, latent decomposition, feature modulation, and adaptive quantization.

\textbf{Recurrent Refinement.}
Early VRC methods obtain scalability through iterative residual coding. Toderici et al.~\citep{toderici2015variable,toderici2017full} introduced recurrent neural codecs in which each iteration encodes additional information to progressively refine the reconstruction. Subsequent studies improved spatially adaptive rate allocation and recurrent coding structures \citep{johnston2018improved,jia2019layered,su2020scalable,islam2021image}. The number of iterations provides a natural rate-control variable, but
higher rates require additional sequential inference, causing computation
and latency to increase with bitrate.
% Early variable-rate learned codecs were primarily based on recurrent architectures. Toderici \textit{et al.}~\citep{toderici2015variable, toderici2017full} proposed recurrent neural networks (RNNs) for LIC, where each iteration progressively encodes additional residual information to refine reconstruction quality. In this framework, bitrate is directly controlled by the number of recurrent iterations, naturally enabling variable-rate coding. Subsequent works~\citep{johnston2018improved, jia2019layered, su2020scalable, islam2021image} further improved network architecture and training stability.

% However, these approaches operate in the pixel-domain residual space and require sequential inference, resulting in increased latency and computational cost at higher bitrates. To alleviate this issue, Yang \textit{et al.}~\citep{yang2021slimmable} proposed slimmable compressive autoencoders (SlimCAEs), where subnetworks with different capacities correspond to different rate levels. Although this design reduces iterative decoding overhead, model complexity still scales with the target bitrate, motivating the development of unified architectures with approximately constant complexity across different rate points.

\textbf{Latent Decomposition.}
A second strategy constructs scalable representations within a single forward pass. Rippel and Bourdev~\citep{rippel2017real} decompose latent representations into ordered components, while multi-scale approaches generate hierarchical representations whose subsets correspond to different reconstruction qualities \citep{nakanishi2018neural,cai2018efficient}. Compared with recurrent refinement, latent decomposition avoids repeatedly executing the complete codec and naturally supports multi-rate or progressive decoding. Nevertheless, the individual representation levels are jointly optimized and can be strongly coupled. Moreover, when rate adaptation is determined by discrete bitplanes, channels, or scales, the available operating points may remain relatively sparse, limiting fine-grained coverage of the R--D curve.
% Another line of work introduces rate scalability through structured decomposition of latent representations within a single forward pass. Rippel \textit{et al.}~\citep{rippel2017real} proposed bitplane decomposition of latent features, where different bitplanes progressively encode finer image details. Similarly, Nakanishi \textit{et al.}~\citep{nakanishi2018neural} and Cai \textit{et al.}~\citep{cai2018efficient} employed multi-scale transforms to generate hierarchical latent representations.

% These methods naturally support progressive transmission and multi-rate decoding without repeated encoding. However, different decomposition levels are often strongly coupled, making joint optimization across all rate points difficult. In addition, because rate adaptation is performed through discrete bitplanes or hierarchical scales, the achievable bitrates are inherently coarse, leading to sparse coverage of the R--D curve.

\textbf{Feature Modulation.}
A more flexible family of methods conditions the transform or latent representation on the desired operating point. Chen and Ma~\citep{chen2020variable} showed that latent representations learned at different bitrates exhibit substantial structural similarity and can be adapted through learned scaling factors. This idea has subsequently been extended by conditioning transform features, normalization parameters, or latent activations on the R--D trade-off parameter \citep{choi2019variable,yang2020variable,cui2021asymmetric,cai2022high,cai2024i2c}. Typically, $\lambda$, a quality index, or another control variable is mapped to a learned embedding that modulates intermediate features.

An important practical property of modulation-based VRC is that the codec topology remains unchanged across operating points. It can therefore provide dense or approximately continuous rate adaptation while maintaining nearly constant computational complexity. The same principle can be extended spatially through quality or importance maps, allowing different regions to operate at different quality levels \citep{song2021variable,ma2021variable,kao2023transformer,liang2023sigvic}. Such spatial control is useful for region-of-interest (ROI) coding \citep{cai2019end}, although determining an optimal quality map introduces an additional rate-allocation problem.

% A more flexible strategy introduces continuous modulation mechanisms into latent representations or transform modules. Chen \textit{et al.}~\citep{chen2020variable} observed that latent representations corresponding to different bitrates exhibit strong similarity, and that bitrate variation can be effectively controlled through latent scaling. Building upon this insight, subsequent works incorporated modulation mechanisms into transform layers, normalization parameters, or latent activations~\citep{yang2020variable, lin2020variable, guo2020variable, choi2019variable, sun2021interpolation, cui2021asymmetric, balle2020nonlinear, gao2021perceptual, cai2022high, cai2024i2c}. Typically, the Lagrangian multiplier $\lambda$, which controls the R--D trade-off, is mapped into a learned embedding that modulates the codec behavior (e.g., convolutional kernels, normalization parameters, or latent activations).

% Compared with recurrent or decomposition-based approaches, feature modulation enables approximately continuous rate control within a single model while maintaining nearly constant complexity across different operating points. Furthermore, several works introduced spatially adaptive quality maps~\citep{cai2024i2c, song2021variable, ma2021variable, kao2023transformer, liang2023sigvic}, enabling fine-grained bit allocation across image regions. This mechanism is particularly promising for region-of-interest (ROI) coding~\citep{cai2019end}, although determining optimal quality maps or ROI regions remains an open problem.

\textbf{Adaptive Quantization.}
Rate adaptation can also be introduced directly at the quantization stage. The basic idea is to scale latent variables before quantization, thereby changing the effective quantization step size and consequently the bitrate. This mechanism is closely related to QP-based control in conventional codecs and can often be integrated without modifying the main transform architecture.

Existing methods differ mainly in the granularity and parameterization of the scaling operation. Global schemes apply a common scale or quantization parameter to a latent tensor \citep{dumas2018autoencoder,zhou2020variable,presta2025stanh}, whereas finer-grained approaches assign different scales to channels or individual latent elements \citep{theis2017lossy,akbari2020learned,akbari2021learned,tong2023qvrf,li2025learnable}. The latter provide greater flexibility in adapting to heterogeneous latent statistics, at the cost of additional parameters or optimization complexity. In either case, the achievable rate range depends on whether the learned representation and entropy model remain well calibrated when the quantization behavior deviates from that seen at individual training points.

Overall, recurrent refinement and latent decomposition provide explicit
scalability but rely on sequential or discrete operating points, whereas
feature modulation and adaptive quantization better preserve a fixed codec
topology and predictable inference complexity. The main remaining challenge
is to maintain stable and monotonic R--D behavior over a sufficiently wide
rate range.
% Inspired by traditional codecs, adaptive quantization remains one of the most fundamental mechanisms for rate control in learned compression (see Fig.~\ref{fig:vrc}d). The core idea is to scale latent representations before quantization, thereby adjusting the effective quantization step size and controlling bitrate.

% Depending on the granularity of scaling, existing methods can be broadly divided into: (i) \emph{global scaling}, where a single scaling factor is applied to all latent variables~\citep{dumas2018autoencoder, zhou2020variable, li2025learnable, presta2025stanh, tong2023qvrf}, and (ii) \emph{latent-wise scaling}, where different latent channels or elements are assigned different quantization factors~\citep{theis2017lossy, gao2021perceptual, akbari2020learned, akbari2021learned, dosovitskiy2019you}. Compared with global scaling, latent-wise strategies provide finer bitrate control and better adaptation to heterogeneous latent statistics.

\subsubsection{Progressive Coding}
\label{sec:progressive}

Progressive coding imposes a stronger constraint than general variable-rate coding. While a VRC model only needs to provide multiple R--D operating points, a progressive codec requires these operating points to be \emph{nested}: a low-rate reconstruction should be obtained from a prefix or ordered subset of the same bitstream, and additional transmitted symbols should progressively refine the reconstruction. This property is particularly useful when the available bandwidth or transmission duration is not known in advance, since the sender can terminate transmission without
re-encoding the image.

Early neural codecs obtain progressiveness through iterative refinement, where additional coding stages successively improve reconstruction quality \citep{toderici2017full,baig2017learning}. Although conceptually simple, such approaches generally require repeated inference as more information is decoded. More recent methods instead construct an ordered representation within a single coding framework, so that progressive reconstruction can be performed by decoding an increasing subset of the latent representation.

\textbf{Structured quantization and latent decomposition.}
One class of approaches explicitly partitions the representation into
ordered coding units. DPICT and its context-based extension encode latents
using trit planes, where successive planes refine the reconstruction
\citep{lee2022dpict,jeon2023context}. Nested quantization constructs
compatible quantization grids across quality levels
\citep{lu2021progressive,lee2026deephq}, while channel- or
resolution-structured representations provide alternative forms of
scalability
\citep{han2020toward,zhang2023exploring}. These methods provide an explicit
progressive structure and well-defined truncation points. However, enforcing
nested representations constrains the optimization of individual operating
points, which may introduce an R--D penalty compared with independently
optimized fixed-rate coding.

\textbf{Importance-aware ordering and masking.}
Another family of methods learns or estimates the relative importance of
latent elements and transmits them accordingly. Importance can be inferred
from latent statistics, learned importance maps, or external feature priors
\citep{lee2022selective,baldassarre2023variable,caron2021emerging}.
ProgDTD~\citep{hojjat2023progdtd} encourages an intrinsic ordering of latent
elements through tail-drop training, allowing a single representation to
support multiple truncation points without substantially changing the codec
architecture. Presta et al.~\citep{presta2025efficient} further use
element-wise importance ranking and variance-aware masking for efficient
progressive decoding. Compared with explicitly layered representations,
masking-based approaches are attractive because progressiveness can be
introduced with relatively limited architectural modification, although
their performance depends on how accurately the ordering reflects the
contribution of each latent element to reconstruction quality.

\textbf{Hierarchical and context-dependent progressiveness.}
Hierarchical representations provide another natural mechanism for
progressive coding. Different latent levels or autoregressive stages can be
decoded successively, producing increasingly refined reconstructions
\citep{cong2026taming,zhang2026autoregressive,zhang2025progressive}.
Such designs can exploit dependencies across representation levels and
achieve strong coding efficiency, but the resulting dependency structure
also reduces decoding independence between stages. This issue becomes
particularly relevant in unreliable networks, where the loss of an early
portion of the representation may affect subsequent decoding. Recent work
therefore combines progressive representation with network-aware training
or loss-resilient coding
\citep{sha2025towards}, linking scalability with communication reliability.

From a deployment perspective, progressive coding enables transmission to
terminate or refine dynamically without re-encoding. Its main challenge is
to provide fine-grained truncation and stable partial reconstruction without
introducing strong representation dependencies or substantial R--D loss.
This trade-off is particularly relevant to bandwidth-limited and unreliable
communication.

\subsubsection{Rate Control}
\label{sec:rate_control}

Variable-rate coding determines the set or range of R--D operating points that a codec can provide, whereas rate control (RC) addresses the inverse problem: given a target bitrate, how should the available coding parameters be selected so that the resulting bitstream satisfies the rate constraint with minimal distortion? This distinction is important in practical systems, where the available bandwidth or storage budget is usually given explicitly rather than represented by an abstract quality parameter. Rate control can be formulated as
\begin{equation}
    \min_{\{\alpha_i\}} D,
    \qquad
    \text{s.t.}\quad R \leq R_{\mathrm{target}},
\end{equation}
where $\{\alpha_i\}$ denotes adjustable coding parameters, such as
quantization scales, modulation factors, or quality indices, and
$R_{\mathrm{target}}$ is the prescribed rate budget. The main challenge is
that the mapping from these parameters to the actual bitrate is
content-dependent and generally cannot be determined analytically.
Existing approaches therefore differ primarily in how this mapping is
estimated and how much additional encoding overhead is introduced.

\textbf{Repeated encoding search.}
The most direct strategy is to encode the image repeatedly with different
coding parameters and search for an operating point close to the target
bitrate. JPEG AI adopts a coarse-to-fine rate-matching procedure, in which
an appropriate model or operating range is first selected and the coding
parameters are subsequently refined
\citep{JPEG-AI,Jia_VRC_2025a}. Jia et al.~\citep{Jia_VRC_2025a} further
describe a two-stage procedure in which a model near the desired rate is
first identified, followed by local estimation of the bitrate--parameter
relationship and iterative refinement toward the target.

Search-based rate control is attractive because it directly evaluates the
actual coded rate and can therefore provide accurate bitrate matching
without requiring a highly accurate predictive model. Its disadvantage is
the need for multiple encoding passes. Since neural encoding itself can be
computationally expensive, repeated evaluation substantially increases
encoding latency and limits its applicability when fast or real-time rate
adaptation is required. Practical search strategies therefore need to
balance the number of evaluations against rate-control accuracy.

\textbf{Bitrate-parameter mapping.}
To reduce repeated encoding, another line of work explicitly models the
relationship between the target bitrate and the codec control parameters.
Shi et al.~\citep{shi2022variable} investigate the relationship between
bitrate and the Lagrange multiplier $\lambda$ and use this relationship for
rate selection. Zhang et al.~\citep{zhang2022rate} introduce a
Rate-Feature-Level (RFL) model that predicts suitable coding parameters
from both the target bitrate and image content. Similarly, Xue and
Zhang~\citep{xue2023lambda} use shallow, high-resolution encoder features to
estimate the mapping from the target rate to $\lambda$. Pan et
al.~\citep{Pan-2025} formulate rate control as a prediction problem over
discrete model and parameter indices, allowing an appropriate configuration
to be selected directly for a given image and target rate.

Compared with repeated search, predictive mapping can substantially reduce
the number of encoding passes and is therefore more suitable for
latency-sensitive applications. The difficulty is shifted to the accuracy
and generalization of the predictor: an inaccurate bitrate-parameter
mapping may lead to rate mismatch, particularly for image content or target
rates that are insufficiently represented during training. Effective rate
control therefore requires the predictor to capture both the global
behavior of the codec and the content-dependent variation of the actual
bitrate.

\textbf{Block-level rate allocation.}
Global rate control assigns a single operating condition to the complete
image and may therefore overlook substantial spatial variation in coding
difficulty. Block-level rate allocation addresses this issue by distributing
a global rate budget across spatial regions according to their local R--D
characteristics. Wang et al.~\citep{Wang-RC-2022} divide an image into
blocks, estimate the bitrate--parameter relationship for individual blocks,
and use a greedy optimization procedure to allocate bits under the global
rate constraint. Because independently estimating such relationships for
all blocks requires repeated encoding, Dong et al.~\citep{Dong-RC-2024}
exploit correlations among neighboring blocks and estimate the behavior of
most blocks from a smaller set of sampled blocks, thereby reducing the
number of required encoding operations.

Block-level control provides finer adaptation to heterogeneous image
content and creates an explicit connection between rate control and spatial
bit allocation. However, it also introduces an additional optimization
layer on top of the neural codec, and its practical value depends on whether
the resulting R--D gain justifies the cost of local rate estimation and
allocation. Similar considerations arise in ROI- and importance-aware
coding, where spatial priorities must ultimately be translated into a
global bit budget.

Overall, rate control introduces an additional requirement beyond the rate
flexibility discussed in Sec.~\ref{sec:variable_rate}: the operating point
must be not only available, but also \emph{predictable and reachable} under
a prescribed rate budget. Repeated search favors accuracy but incurs
encoding overhead, predictive mapping reduces computation but depends on
generalization, and block-level allocation provides finer spatial control
at the cost of additional optimization. A practical RC scheme therefore
needs to balance rate accuracy, encoding complexity, and R--D performance,
while remaining compatible with the variable-rate mechanism of the
underlying codec.

\subsubsection{Discussion and Perspective}

Practical rate adaptation should be evaluated beyond R--D performance
alone. A deployable codec should provide sufficiently wide and smooth rate
coverage, accurate target-rate matching, limited control overhead, and
predictable inference complexity. Modulation-based VRC and adaptive
quantization are particularly attractive in this respect because they retain
a fixed codec topology while providing dense operating-point adaptation
\citep{duan2023qarv,jia2025towards}.

A remaining challenge is the content dependence of the mapping between
control parameters and actual bitrate. Spatial quality maps, ROI coding,
and block-level allocation provide finer adaptation to heterogeneous image
content \citep{ma2021variable,kao2023transformer,Wang-RC-2022,
Dong-RC-2024}, but local decisions must ultimately satisfy a global bitrate
constraint. Joint modeling of global rate budgets and spatial importance is
therefore a promising direction.

More generally, rate adaptation interacts with other deployment
requirements. Low-precision inference should preserve the behavior of the
rate-control mechanism, while robustness and transmission resilience may
change both the available bitrate and its preferred allocation. Extending
these principles to video further introduces temporal dependencies, buffer
constraints, and sequential allocation, as discussed in Sec.~\ref{sec:conclusion}.

\begin{figure*}[t]
\centering
\includegraphics[width=0.95\linewidth]{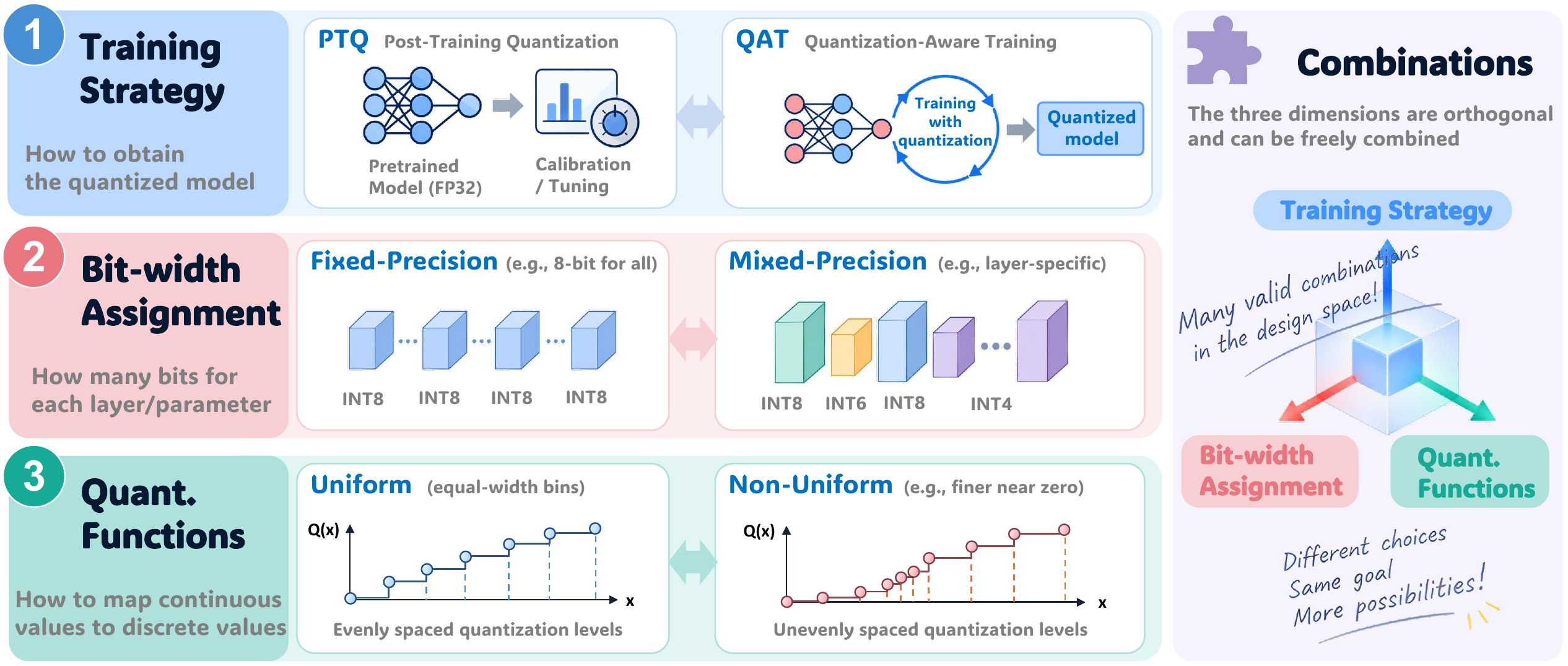}
\caption{Three complementary design dimensions of model quantization for LIC: optimization strategy (QAT or PTQ), bit-width assignment (fixed or mixed precision), and quantization function (uniform or non-uniform). These dimensions can be combined according to R--D sensitivity and hardware constraints, while cross-platform consistency introduces an additional codec-level requirement discussed in Sec.~\ref{sec:mixed_precision}.}
\label{fig:quant}
\end{figure*}

\subsection{Model Quantization and Cross-Platform Consistency}
\label{sec:quantization_consistency}

As illustrated in Fig. \ref{fig:quant}, model quantization plays a particularly important role in practical LIC.
Unlike the signal quantization discussed in Sec.~\ref{sec:core_framework}, which
discretizes latent representations for entropy coding, model quantization
reduces the numerical precision of network weights and activations. It
serves two closely related deployment objectives. First, low-precision
arithmetic reduces model size, memory traffic, and computational cost, and
facilitates execution on integer-oriented accelerators. Second, deterministic
numerical behavior is required by a codec whose encoder and decoder may run
on different hardware and software platforms.

The latter requirement makes quantization in LIC fundamentally different
from that in conventional recognition networks. In a classification model,
small numerical differences usually result only in small changes in the
output scores. In a learned codec, however, the entropy decoder relies on
probability parameters reproduced at both the encoder and decoder. Small
floating-point discrepancies in this shared probability-estimation path may
produce different probability tables and eventually desynchronize entropy
decoding. Model quantization is therefore not only a model-compression
technique, but also an important mechanism for codec interoperability.

Early work by Ball\'e et al.~\citep{balle2018integer} demonstrated that
learned compression can be implemented using integer arithmetic, providing
a foundation for deterministic execution. Subsequent studies have explored
different quantization strategies for balancing R--D performance,
computational efficiency, and numerical consistency. We review these methods
from three perspectives: the optimization procedure, the precision
granularity, and the level of cross-platform consistency required by the
codec.

\subsubsection{Quantization-Aware Training and Post-Training Quantization}
\label{sec:qat_ptq}

From the optimization perspective, model quantization methods can be broadly
divided into quantization-aware training (QAT) and post-training
quantization (PTQ). The two approaches differ mainly in whether the model is
allowed to adapt to low-precision arithmetic during training.

\textbf{Quantization-aware training.}
QAT simulates low-precision arithmetic during optimization so that model
parameters can adapt to the resulting quantization error. Early studies
first quantized convolutional weights
\citep{sun2020end,sun2021learned} and later extended low-precision
computation to both weights and activations
\citep{sun2021end}. Compared with directly quantizing a pretrained network,
QAT generally provides greater flexibility in compensating for numerical
errors, but requires access to the training data and an additional training
stage.

LIC also introduces compression-specific difficulties that are not fully
captured by generic neural-network quantization. The transform and entropy
models affect not only reconstruction features but also latent statistics
and probability estimation. Consequently, quantization error may influence
both distortion and coding rate. Hong et al.~\citep{hong2020efficient}
developed fixed-point inference with adaptive numerical ranges for efficient
neural image decoding. Q-LIC~\citep{sun2022q} reduces quantization error by
splitting channels with large dynamic ranges, while IQ-LIC
\citep{jeon2023integer} introduces quantization-related supervision to
improve integer inference. QAT has also been combined with structured
pruning \citep{hossain2024structured} and variable-rate coding
\citep{ye2025variable}, illustrating that model precision must often be
optimized together with other deployment constraints.

The main advantage of QAT is therefore its ability to preserve R--D
performance under a prescribed numerical format. Its main limitation is
deployment flexibility: every change in precision configuration may require
additional optimization, which becomes costly when multiple hardware
targets or codec configurations must be supported.

\textbf{Post-training quantization.}
PTQ converts a pretrained floating-point codec to low precision without
full retraining. This substantially reduces deployment cost and makes it
possible to quantize existing models after model development. Generic PTQ
methods typically minimize the discrepancy between floating-point and
quantized weights or activations
\citep{nagel2021white,li2021brecq}, but such criteria are not necessarily
aligned with the objective of image compression.

He et al.~\citep{he2022post} adapt PTQ to learned image compression and
combine low-precision inference with deterministic entropy coding for
cross-platform deployment. RDO-PTQ~\citep{shi2023rate} further shows that
minimizing tensor-level quantization error does not necessarily minimize
the degradation in compression performance. Instead, the quantization
configuration should be optimized according to its effect on the R--D
objective. This observation is particularly relevant to LIC because
different modules affect rate and distortion in substantially different
ways.

QAT and PTQ therefore provide complementary deployment strategies. QAT
offers stronger adaptation to low precision and is attractive when the
target hardware configuration is known during model development. PTQ
preserves the original training pipeline and is more suitable for converting
existing codecs or supporting multiple deployment targets. In both cases,
however, determining \emph{where} and \emph{how aggressively} to quantize
the codec is as important as choosing the optimization procedure itself.

\subsubsection{Precision Granularity and Mixed-Precision Quantization}
\label{sec:mixed_precision}

A second design dimension concerns the granularity of numerical precision.
The simplest strategy applies a uniform precision, such as INT8, throughout
the network. Uniform quantization is straightforward to implement and maps
well to commonly available low-precision hardware. However, different
codec modules, layers, and channels can exhibit substantially different
sensitivities to numerical error. Applying the same bit width everywhere
may therefore allocate excessive precision to insensitive components while
over-quantizing sensitive ones.

Mixed-precision quantization (MPQ) addresses this issue by assigning
different numerical precisions according to their impact on compression
performance. FMPQ~\citep{hossain2024flexible} and mixed-precision
PTQ~\citep{yu2025mixed} evaluate the sensitivity of individual components
to quantization and allocate precision accordingly. Other approaches
perform finer-grained selection over candidate bit widths
\citep{yang2025subset}, while DynaQuant~\citep{bao2026dynaquant} introduces
dynamic routing among different precision configurations. These methods
enlarge the design space from selecting a single bit width to jointly
optimizing precision allocation and R--D performance.

Another line of work modifies the numerical distribution before
quantization. Long-tailed or highly non-uniform activations may lead to poor
utilization of a uniformly spaced quantization range. Distribution
balancing or transformation can reduce this mismatch and improve PTQ
performance. For example, Yu et al.~\citep{yu2025activation} adapt weight
and activation distributions to reduce quantization error before
low-precision inference.

From an algorithmic perspective, fine-grained MPQ provides an attractive
performance--precision trade-off. From a deployment perspective, however,
the effective benefit depends on hardware support. A theoretically favorable
combination of, for example, 4-, 6-, and 8-bit operations does not
necessarily translate into lower latency if the target accelerator executes
these data types inefficiently or requires frequent precision conversion.
Thus, precision allocation should ultimately be evaluated using both R--D
sensitivity and the operator support of the target hardware, rather than
bit width alone.

\subsubsection{Cross-Platform Consistency}
\label{sec:cross_platform}

For a codec, low-precision inference and cross-platform consistency are
related but not identical objectives. Quantizing the complete network may
improve hardware efficiency, but deterministic entropy decoding only
requires numerical agreement in the computation that is shared between the
encoder and decoder and affects symbol probabilities. This distinction
provides an opportunity to achieve interoperability without unnecessarily
quantizing every component of the codec.

Koyuncu et al.~\citep{koyuncu2022device} identify the learned probability
model as a primary source of device-dependent entropy-decoding mismatch.
Their results show that, for the studied still-image codecs, quantizing the
shared entropy-related modules is sufficient to obtain consistent entropy
decoding while leaving other parts of the transform in floating point.
Subsequent work further quantizes decoder computation to provide
deterministic reconstruction
\citep{koyuncu2024quantized}. These results suggest two different levels of
numerical consistency: \emph{entropy-decoding consistency}, which guarantees that
the transmitted bitstream is decoded into the same latent symbols, and
\emph{reconstruction consistency}, which additionally requires the decoded
pixels to be numerically reproducible across platforms.

Alternative mechanisms attempt to tolerate or remove the source of
numerical mismatch. Pang et al.~\citep{pang2024towards} introduce auxiliary
safeguarding information that allows the decoder to tolerate limited
numerical deviations, although the additional information increases the
bitstream overhead. Rather than forcing identical learned probability
estimation, Yang et al.~\citep{yang2025learned} explore cross-platform
entropy coding based on chain coding, providing another route toward
reproducible decoding.

This distinction between full-network quantization and selective
deterministic computation is important in practice. If interoperability is
the primary goal, quantizing only the modules that affect entropy decoding
can preserve most of the floating-point transform while minimizing the
R--D penalty. If low-power or accelerator-oriented inference is also
required, broader or full-network quantization becomes necessary. Practical
codec design should therefore determine the required level of numerical
consistency before deciding the scope of quantization.

\subsubsection{Discussion and Perspective}
\label{sec:quant_discussion}

Model quantization in learned image coding serves two related but distinct
deployment objectives: efficient low-precision inference and deterministic
codec operation. The former generally favors broader quantization of
transforms and entropy models, whereas the latter may require deterministic
computation only along the encoder--decoder shared probability-estimation
path. Selective quantization can therefore provide cross-platform
entropy-decoding consistency with limited R--D degradation, while full or
mixed-precision quantization further targets runtime and energy efficiency.

The practical benefit of a numerical format also depends strongly on
hardware support. Although mixed-precision designs can reduce quantization
error, accelerators typically provide efficient kernels for only a limited
set of data types. Consequently, precision selection should consider
operator availability, memory behavior, and execution efficiency together
with R--D sensitivity. This issue becomes more challenging in predictive
video coding, where small numerical mismatches may accumulate along the
prediction chain, making efficient deterministic decoding an important
open problem \citep{van2024mobilenvc,conceiccao2025cross}.

\subsection{Robustness and Communication Resilience}
\label{sec:robustness}

Beyond nominal R--D performance, practical LIC must remain reliable when
the input distribution, numerical conditions, or communication environment
deviates from the ideal setting used during training and evaluation. This
requirement is particularly important for learned codecs because nonlinear
transforms and data-driven entropy models may respond unpredictably to
small perturbations. A perturbed input may affect not only reconstruction
quality, but also latent statistics, probability estimation, and ultimately
the coded bitrate.

In this survey, robustness is therefore considered as a system-level
property rather than only an adversarial-learning problem. It includes
sensitivity to adversarial perturbations and backdoor attacks, but also
repeated compression, benign distribution shifts, packet loss or bitstream
corruption, and the propagation of compression errors to downstream vision
tasks. These failure modes arise at different stages of the coding and
communication pipeline and expose different trade-offs between compression
efficiency and reliability.

\begin{figure}[t]
\centering
\includegraphics[width=1.0\linewidth]{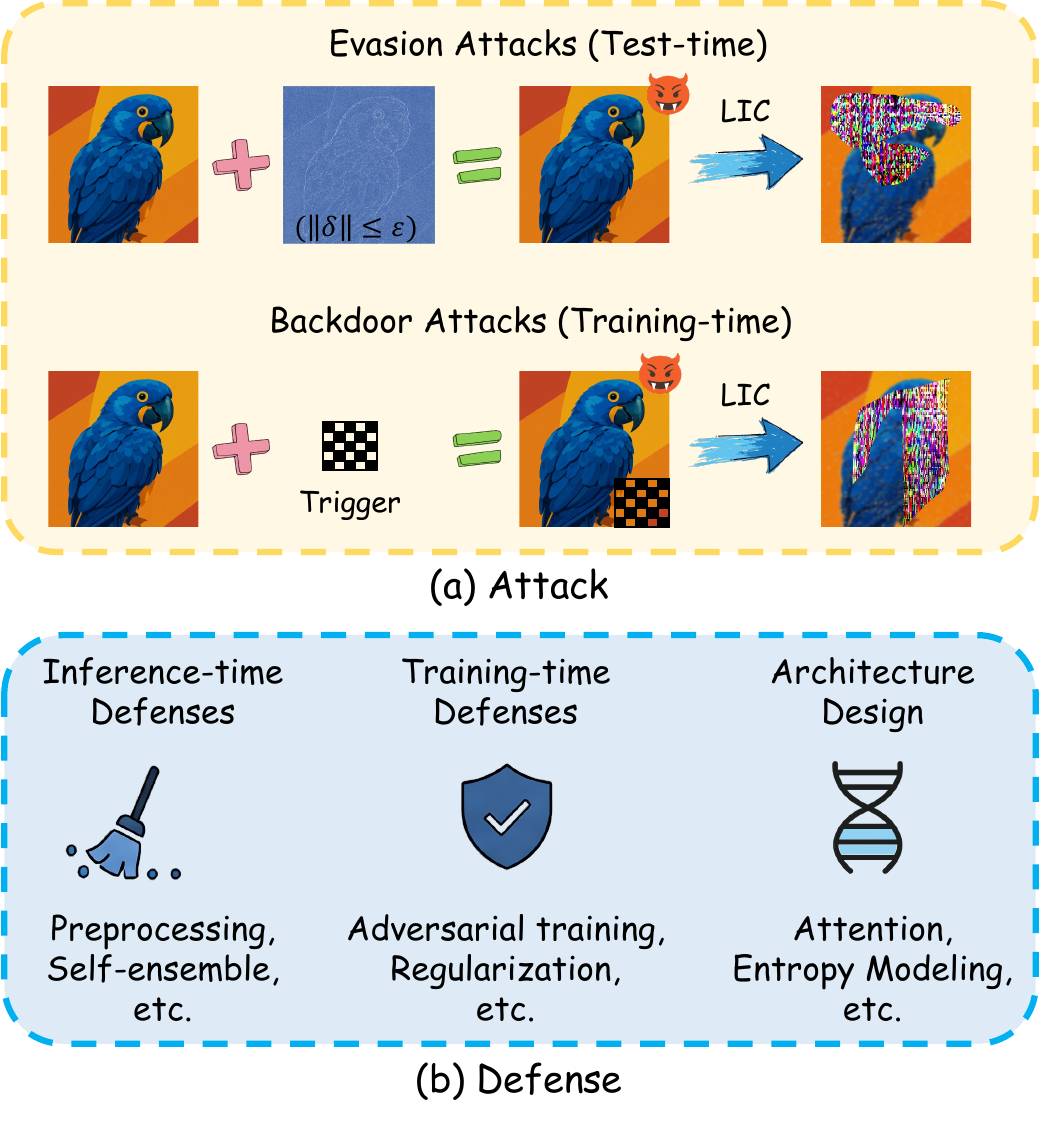}
\caption{Illustration of attacks and defenses for LIC models. Attacks include test-time evasion perturbations and training-time backdoor triggers, while defenses can be introduced through architecture design, robust training, and inference-time protection.}
\label{fig:attack}
\end{figure}

\subsubsection{Adversarial Vulnerability}
\label{sec:robust_attack}

Adversarial robustness has received increasing attention as learned codecs
become part of larger visual-processing pipelines. Given an input image
${x}$, an adversarial example can be written as
\begin{equation}
    {x}^{*} = {x} + {\delta},
    \qquad
    \|{\delta}\| \leq \epsilon,
\end{equation}
where the perturbation ${\delta}$ is constrained to remain small
while causing undesirable codec behavior. Unlike attacks on conventional
recognition networks, the objective in LIC may target reconstruction
distortion, coding rate, or both.

\textbf{Evasion attacks.}
In the white-box setting, gradients through the learned codec can be used to
directly optimize perturbations against compression-specific objectives.
Chen and Ma~\citep{chen2023toward} introduce a distortion-oriented attack
that substantially degrades reconstruction quality under bounded input
perturbations. Subsequent work explores more efficient adversarial
optimization \citep{kurihara2025efficient}, bitstream-level failure modes
\citep{madden2025bitstream}, transform-domain attacks
\citep{kalmykov2026t}, and objectives that explicitly manipulate the
R--D behavior of the codec \citep{wu2025adversarial}.

In black-box settings, perturbations generated on surrogate models may
transfer to unseen compression architectures
\citep{papernot2016transferability,liu2017delving}. Such transferability is
particularly relevant to deployment because an attacker may not have access
to the exact codec parameters. Recent work further shows that perturbations
introduced at the compression stage can propagate to downstream machine
vision systems \citep{sui2024transferable}, extending the impact from
reconstruction quality to task-level reliability.

\textbf{Backdoor attacks.}
A different threat arises during model training or fine-tuning. Backdoor
attacks insert hidden triggers into part of the training data so that the
codec behaves normally on clean inputs but exhibits abnormal behavior when
the trigger is present. Frequency-domain triggers have been studied for
learned image compression
\citep{yu2023backdoor,yu2024robust}, while compression-resistant trigger
designs aim to preserve malicious behavior after image coding
\citep{xue2023compression}. These attacks are especially relevant when
models are trained, adapted, or distributed using data and checkpoints from
untrusted sources.

From a deployment perspective, adversarial vulnerability is important not
only because a codec may be intentionally attacked, but also because these
studies reveal regions of numerical and statistical instability in learned
representations. They therefore provide useful stress tests for evaluating
codec reliability beyond conventional clean-image benchmarks.

\subsubsection{Robustification Strategies}
\label{sec:robust_defense}

Existing defenses can be broadly categorized according to where robustness
is introduced: in the codec architecture, during training, or at inference
time.

\textbf{Architecture-level robustness.}
The structure of the entropy model can strongly affect sensitivity to input
perturbations. Highly expressive autoregressive models improve probability
estimation under nominal conditions, but they also introduce stronger
dependencies among coded symbols. Studies of manipulation attacks suggest
that simplifying or stabilizing these dependencies can improve robustness
\citep{liu2022malice,liu2023manipulation}. This reveals a recurring
trade-off in practical LIC: stronger statistical modeling may improve R--D
efficiency while increasing sensitivity to distribution mismatch or error
propagation.

\textbf{Training-time robustification.}
Adversarial training incorporates perturbed samples during optimization so
that the codec learns to maintain stable reconstruction under bounded input
changes
\citep{chen2023toward,madry2018towards}. Compression-specific variants can
combine robust optimization with representation consistency or knowledge
transfer. For example, Cao et al.~\citep{cao2024enhancing} integrate
knowledge distillation with adversarial training, while successive-coding
stability can be improved by constraining latent or reconstruction
consistency across repeated compression cycles
\citep{kim2020instability}. Training-based methods generally provide stronger
adaptation to known perturbation models, but they increase training cost and
may introduce a penalty on clean-image R--D performance.

\textbf{Inference-time protection.}
Robustness can also be improved without modifying the original training
procedure. Representative strategies include input preprocessing,
self-ensemble, randomized transformations, and test-time filtering
\citep{chen2023toward,song2024training}. Compression itself can also act as
a preprocessing defense by removing high-frequency adversarial noise
\citep{jia2019comdefend}. Such approaches are attractive because they can be
applied to existing codecs, but their effectiveness may depend strongly on
the attack model and can degrade against adaptive perturbations.

These three classes of methods provide different practical trade-offs.
Architecture-level design changes the codec itself, training-time defenses
increase development cost but can provide stronger robustness, and
inference-time defenses offer greater flexibility at the expense of less
predictable protection.

\subsubsection{Practical Robustness and Communication Resilience}
\label{sec:practical_robustness}

Adversarial attacks represent only one form of non-ideal input. In deployed
compression systems, more common reliability problems arise from repeated
processing, content variation, and imperfect communication channels.

\textbf{Repeated compression.}
Images are often compressed multiple times as they move across storage,
editing, communication, and social-media pipelines. Learned codecs may
exhibit accumulated reconstruction drift under successive
encoding--decoding cycles
\citep{kim2020instability}. Unlike a single-pass R--D evaluation, repeated
compression exposes whether reconstruction errors alter the latent
distribution in a way that is amplified by subsequent coding stages.
Robustness to such repeated processing is therefore relevant to practical
interoperability and long-term media reuse.

\textbf{Transmission errors.}
A more direct communication challenge is packet loss or bitstream
corruption. Conventional coding systems typically rely on channel-side
protection, such as forward error correction (FEC)
\citep{mizuochi2006recent}. Learned codecs provide an additional possibility:
the representation and decoder can themselves be optimized for incomplete
or corrupted observations.

LossResilientLIC~\citep{sha2025towards} incorporates network perturbations
during training and constructs representations that remain decodable when
part of the transmitted information is lost. ResiComp
\citep{wang2025resicomp} similarly studies loss-resilient representations
through masked visual-token modeling, while progressive probability models
have been explored for streaming over unreliable channels
\citep{zhang2025learned}. These approaches shift part of the reliability
problem from channel coding to source representation.

However, robustness to packet loss is closely coupled with entropy modeling.
Autoregressive dependencies can improve coding efficiency under error-free
transmission, but missing symbols may invalidate the context required for
subsequent decoding and amplify error propagation. More independent or
hierarchical representations can reduce this dependency, but often at a
cost in nominal R--D performance. Communication resilience therefore
introduces an explicit performance--reliability trade-off.

\textbf{Interaction with downstream tasks.}
Compression is increasingly used as a front-end component of recognition,
detection, and other machine-vision pipelines. In such systems, codec
robustness cannot be evaluated solely from reconstructed PSNR or perceptual
quality. Compression may suppress adversarial noise and improve downstream
stability \citep{jia2019comdefend}, but artifacts or manipulated coding
behavior may also propagate to later models
\citep{liu2024backdoor,sui2024transferable}. The relevant system-level
criterion is therefore whether the compressed representation preserves the
information required by the downstream task under realistic perturbations.

\subsubsection{Discussion and Perspective}
\label{sec:robust_discussion}

Robustness in learned image coding should be viewed as a system-level
reliability problem rather than an isolated adversarial-learning issue.
Practical evaluation should therefore consider not only performance under
perturbations or packet loss, but also the clean-condition R--D cost
introduced by the corresponding robustness mechanism. In particular,
stronger protection may require additional training, redundancy, simplified
entropy dependencies, or error-resilient representations, each of which can
affect bitrate, distortion, or complexity.

A remaining challenge is the lack of common evaluation protocols across
different robustness settings. Perturbation budgets, attack objectives,
packetization strategies, and channel models vary substantially across
studies, making direct comparison difficult. Standardized test conditions
that jointly report clean performance and degradation under specified
impairments would therefore be valuable for deployment-oriented LIC.
Overall, the desired codec should remain stable under input variation and
degrade gracefully when transmitted information is incomplete.

\begin{figure*}[t]
\centering
\includegraphics[width=0.95\linewidth]{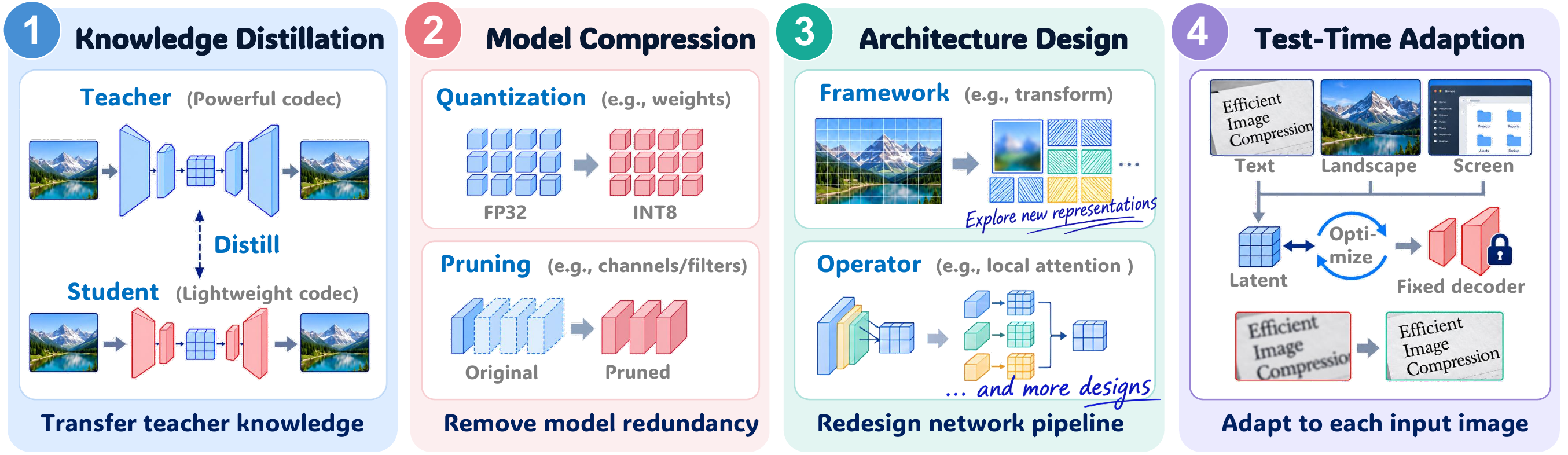}
\caption{A taxonomy of strategies for lightweight and hardware-aware deployment in LIC, including knowledge distillation, model compression, architecture design, and test-time adaptation.}
\label{fig:deployment}
\end{figure*}

\subsection{Lightweight and Hardware-Aware Deployment}
\label{sec:lightweight}

Recent LIC models have achieved substantial improvements in compression
efficiency, with several methods approaching or surpassing conventional
coding anchors under commonly used R--D benchmarks. These gains, however,
are often accompanied by increasingly large neural transforms, complex
entropy models, and substantial computational overhead. As illustrated in
Table~\ref{tab:lic_overview}, many high-performance codecs require several
hundred or even more than one thousand KMACs/pixel, which remains
prohibitive for resource-constrained devices.

Practical deployment therefore requires a broader notion of efficiency than
model size or multiply--accumulate operations alone. Latency, memory
footprint, parallelism, operator support, and entropy-coding overhead can all
affect the actual runtime of a codec. Moreover, encoder and decoder
complexities need not be symmetric: in many applications, additional
computation can be tolerated at the encoder if it enables a substantially
lighter decoder. Existing work addresses these constraints through knowledge
distillation, model compression, architecture-level redesign, and
instance-specific adaptation, as summarized in Fig. \ref{fig:deployment}.

\subsubsection{Knowledge Distillation}
\label{sec:distillation}

Knowledge distillation (KD) transfers information from a high-capacity
teacher model to a smaller student network
\citep{gou2021knowledge}. In learned image compression, this objective is
more involved than matching class predictions in recognition networks,
because the student must reproduce not only reconstruction behavior but also
compact latent representations and accurate probability models.

Early approaches use reconstruction- or entropy-related supervision to
transfer coding behavior from the teacher
\citep{fu2024fast}. More recent studies introduce finer-grained supervision
at intermediate features or latent representations
\citep{allemand2025efficient}. Chen et al.~\citep{chen2025knowledge} further
perform multi-level distillation across network blocks, allowing the
lightweight student to inherit representation and reconstruction capability
from a larger model.

Rather than explicitly defining distillation losses, EVC
\citep{wang2023evc} adopts a mask-decay strategy that progressively reduces
the effective model capacity during training, thereby transferring the
behavior of a larger network into a compact architecture. CSLIC
\citep{wang2026distilling} further combines KD with neural architecture
search to derive multiple encoder--decoder configurations at different
complexity levels.

From a deployment perspective, KD is attractive because complexity reduction
is achieved mainly during model development and does not introduce
additional inference-time modules. Its effectiveness, however, depends on
how much of the teacher's coding capability can be retained under a
substantially smaller transform or entropy model. Distilling only
reconstruction features may therefore be insufficient when the dominant
performance loss originates from probability modeling or latent
representation capacity.

\subsubsection{Model Compression: Quantization and Pruning}
\label{sec:model_compression}

Model compression provides a second route to lightweight LIC by reducing
the numerical or structural redundancy of a trained network. Quantization
reduces the precision of weights and activations and can substantially
improve memory and arithmetic efficiency
\citep{gholami2022survey}. Because its implications for R--D performance
and cross-platform consistency have been discussed in detail in
Sec.~\ref{sec:quantization_consistency}, we focus here primarily on its role
as a computational optimization.

Pruning instead removes redundant parameters or computational structures
from the network
\citep{zhu2017prune}. For LIC, structured pruning is generally more relevant
to practical deployment than unstructured sparsity because removing complete
channels or filters can translate more directly into reduced computation.
Liao et al.~\citep{liao2022efficient} apply structured pruning specifically
to the decoder, demonstrating that substantial complexity reduction can be
achieved while retaining competitive reconstruction performance.
Quantization and pruning can also be combined
\citep{hossain2024structured}, jointly reducing arithmetic precision and
network capacity.

The practical effectiveness of model compression ultimately depends on
whether the resulting sparsity or numerical precision is efficiently
supported by the target platform. Parameter reduction alone does not
necessarily imply proportional runtime reduction. Consequently,
hardware-oriented compression should consider deployable tensor shapes,
supported numerical formats, and efficient operators rather than only
nominal parameter counts.

\subsubsection{Architecture-Level Design}
\label{sec:efficient_architecture}

A more direct strategy is to design the codec for efficiency from the
beginning rather than compressing an already expensive model. This includes
reducing transform complexity, simplifying entropy dependencies, improving
parallelism, and allocating computation asymmetrically between the encoder
and decoder. Such architecture-level optimization is particularly important
because computational structures that provide strong R--D performance on
GPUs may not map efficiently to mobile or edge accelerators.

\textbf{Efficient entropy modeling.}
Entropy models are a major source of practical decoding overhead.
Autoregressive contexts improve probability estimation by exploiting
dependencies among latent symbols, but their sequential nature can restrict
parallelism and increase latency. Ali et al.~\citep{ali2023towards} explore
efficient learned compression without conventional autoregressive
dependencies, illustrating that part of the R--D gain from stronger context
modeling can be traded for substantially improved execution efficiency.
More generally, reducing the number and granularity of sequential context
steps is an important design dimension when decoding throughput is a primary
constraint.

\textbf{Asymmetric codec design.}
Encoder and decoder complexity need not be balanced. In many storage,
delivery, or content-distribution scenarios, an image is encoded once but
decoded repeatedly on heterogeneous client devices. This motivates shifting
computation toward the encoder while maintaining a lightweight decoder.
AsymLLIC~\citep{wang2024asymllic}, for example, explicitly adopts an
asymmetric encoder--decoder architecture to reduce decoding complexity.
ShallowNTC~\citep{yang2023computationally} pushes this principle further by
using a shallow, nearly linear synthesis transform, achieving performance
competitive with BPG at less than 50 KMACs/pixel decoding complexity.

\textbf{Efficient operators and dynamic computation.}
Another direction replaces expensive neural operations with more
hardware-friendly alternatives. Lightweight attention
\citep{he2024towards,chen2026lightweight}, efficient convolutions
\citep{li2022lightweight}, and shift-based processing have all been explored
to reduce redundant computation. ShiftLIC~\citep{bao2025shiftlic}, for
example, replaces computationally expensive spatial operations with
spatial--channel shifts and small convolutions. ULIC
\citep{dong2023ulic} similarly targets highly constrained wearable-device
settings through an explicitly lightweight codec design.

FastNIC~\citep{zhang2025learning} jointly considers transform and prior
complexity through switchable entropy priors, achieving BPG-level or better
performance on Kodak with encoding and decoding complexities below
approximately 12 and 10 KMACs/pixel, respectively. Beyond fixed
architectures, AdaNIC~\citep{tao2023adanic} uses dynamic transform routing
to adapt computational paths according to image content. Such conditional
computation provides another dimension of flexibility: instead of assigning
the same complexity to every image, the codec can allocate computation
according to coding difficulty.

These examples suggest that lightweight LIC increasingly requires joint
optimization of the transform and entropy-model execution paths. Reducing
the complexity of only one component may leave another component as the
runtime bottleneck, while nominal MAC reduction may provide limited benefit
if the resulting operators are poorly supported by the target hardware.

\subsubsection{Test-Time Adaptation}
\label{sec:tta}

Test-time adaptation (TTA) provides a different approach to the
performance--complexity trade-off. Rather than permanently increasing model
capacity, TTA spends additional computation or transmits a small number of
instance-specific parameters to adapt a shared codec to the current image
or domain
\citep{chen2022contrastive,liang2025comprehensive}. In learned compression,
this strategy is closely related to the amortization gap of VAE-based
models, where a fixed encoder may not produce the optimal latent
representation for every input
\citep{yang2020improving}.

\textbf{Latent refinement.}
Encoder-side adaptation directly optimizes the latent representation for the
current image under the R--D objective. Campos et al.
\citep{campos2019content} refine latent variables through iterative
optimization, while Yang et al.~\citep{yang2020improving} jointly optimize
latent representations and associated side information to reduce the
amortization gap and discretization error. These approaches can improve
compression performance without increasing decoder complexity or model
size, but require additional optimization iterations during encoding.

\textbf{Decoder-side adaptation.}
Another family of methods adapts part of the decoder and transmits the
corresponding parameter updates together with the bitstream. Instance-adaptive
compression~\citep{van2021overfitting} optimizes decoder and entropy-model
parameters for individual inputs, while related work restricts adaptation to
a smaller subset of parameters, such as biases or multiplicative
coefficients
\citep{lam2020efficient,zou2021adaptation}. The additional adaptation can
improve coding efficiency, but its benefit must compensate for both the
optimization cost and the bits required to signal the parameter updates.

Parameter-efficient transfer learning provides a more lightweight
alternative. Inspired by techniques such as LoRA
\citep{hu2022lora}, recent compression methods introduce compact adapters or
low-rank updates instead of modifying the full decoder
\citep{shen2023dec,tsubota2023universal,lv2023dynamic}. Such modules are
particularly attractive for adapting a common codec to different content
domains. Chen et al.~\citep{chen2025test} further study test-time adaptation
under distribution shifts and regularize adaptation of the probability
model to improve cross-domain compression.

TTA therefore exposes a different form of resource allocation from
lightweight architecture design. A compact shared model can recover part of
the performance of a larger model by spending additional encoder-side
computation or transmitting instance-specific information. This trade-off
is attractive when decoding complexity and persistent model size are
strictly constrained but additional encoding effort is acceptable.

\subsubsection{Discussion and Perspective}
\label{sec:lightweight_discussion}

Lightweight LIC should be evaluated as a system-level
performance--complexity problem. Parameter count and KMACs/pixel provide
useful architecture-level proxies, but they do not fully characterize
deployment efficiency. Sequential entropy decoding, memory movement,
intermediate feature size, operator availability, and hardware utilization
can cause models with similar arithmetic complexity to exhibit substantially
different runtime behavior. Actual latency, memory consumption, and target
hardware should therefore be considered together with algorithmic
complexity.

More broadly, reducing model complexity is useful only when the resulting
computation maps efficiently to the target platform. Distillation, pruning,
quantization, architecture-level redesign, and test-time adaptation provide
complementary ways to trade model capacity, numerical precision, and
encoder-side computation for efficiency. Practical LIC should therefore
optimize the R--D--complexity trade-off jointly with entropy-model
dependencies, numerical precision, memory behavior, parallelism, and
hardware-supported operators. The TinyLIC benchmark in Sec.~4 follows this
perspective by exposing multiple computational budgets under a common codec
topology.

\subsection{Standardization and Interoperability}
\label{sec:standardization}

Standardization represents a critical step in the transition from
high-performance learned compression models to deployable codecs. A
practical coding standard must define not only compression performance, but
also interoperable bitstreams, decoder behavior, operating profiles,
complexity configurations, test conditions, and conformance requirements.
These issues are particularly important for LIC, where model architecture,
numerical precision, entropy-model execution, and hardware implementation
can all affect reproducibility across platforms.

Recent standardization activities indicate that learned image coding is
moving beyond isolated research prototypes toward interoperable coding
systems \citep{liu2024guest,gao2025standards}. Two representative efforts are
JPEG AI and IEEE 1857.11. Although their specific architectures and
evaluation methodologies differ, both retain the basic learned
transform-coding paradigm reviewed in Sec.~\ref{sec:core_framework}: an analysis
transform maps the input image to quantized latent representations, which
are entropy coded and reconstructed by a learned decoder. More importantly,
both standards extend this basic formulation with explicit considerations
of complexity, functionality, and implementation.

\subsubsection{JPEG AI}
\label{sec:jpeg_ai}

JPEG AI is an international standard for end-to-end learning-based image
coding developed by the ISO/IEC Joint Photographic Experts Group
\citep{JPEG-AI,alshina2024jpeg}. The activity was initiated in 2019 and has
subsequently evolved into the ISO/IEC 6048 / ITU-T T.840 family of
standards. Rather than specifying only a high-performance neural model,
JPEG AI addresses a broader set of system requirements, including
multi-complexity decoding, rate adaptation, progressive and partial
decoding, compressed-domain processing, and interoperable implementation
\citep{esenlik2025overview}.

\textbf{Multi-complexity coding architecture.}
An important characteristic of JPEG AI is the availability of decoder
configurations spanning substantially different computational budgets.
Reported configurations range from approximately 8 to more than
200 KMACs/pixel, allowing the same standardization framework to target
different hardware classes \citep{esenlik2025overview}. A configuration of
approximately 23 KMACs/pixel has also been demonstrated in mobile-oriented
implementations, illustrating the importance of explicitly considering
decoder complexity when defining a practical learned codec.

The coding framework also adopts structured color processing to balance
compression efficiency and memory consumption. Input images are converted
to a YUV representation, where luminance is encoded using a dedicated
neural path and chrominance components are coded conditionally on the
reconstructed luminance. Such asymmetric processing exploits inter-channel
correlation while reducing the computational and memory requirements of
chrominance coding \citep{esenlik2025overview}.

\textbf{Extended coding functionalities.}
JPEG AI incorporates several functionalities beyond conventional
single-point image reconstruction. Progressive decoding is enabled by
organizing transmitted representations according to their importance, so
that reconstruction quality can improve as additional information becomes
available. Spatially partial decoding is supported through tiled
representations, allowing selected image regions to be reconstructed without
processing the complete image. These capabilities connect standardization
with the variable-rate and progressive coding mechanisms discussed in
Sec.~\ref{sec:rate_flex}.

The standard also considers processing directly from compressed
representations. Representative functionality includes image
classification, super-resolution, and denoising
\citep{JPEG-AI,esenlik2025overview}. This direction extends the role of the
coded representation beyond human-view reconstruction and provides an
early standardized interface between image compression and downstream
visual processing. Recent studies have further investigated
compressed-domain analysis using JPEG AI representations
\citep{alkhateeb2024jpeg}.

\textbf{Evaluation and implementation.}
JPEG AI places substantial emphasis on perceptual quality and reproducible
evaluation. Its Common Test and Training Conditions (CTTC) specify common
datasets, coding conditions, and perceptual quality measures. The evaluation
framework includes metrics such as MS-SSIM, IW-SSIM, VMAF, VIF, NLPD, and
FSIM, reflecting a broader quality criterion than conventional
PSNR-oriented evaluation \citep{JPEG-AI,esenlik2025overview}.

Implementation is also explicitly considered. In representative
implementations, entropy decoding can be executed on CPUs to preserve
bit-exact behavior, while computationally intensive neural transforms are
mapped to GPUs or NPUs. Such heterogeneous execution illustrates a
practical principle that also appears throughout this survey: the most
efficient implementation need not execute the complete codec using a
single numerical format or hardware backend. Instead, deterministic and
computationally intensive components can be assigned to different
processors according to their requirements.

JPEG AI has consequently become an important platform for studying
deployment-oriented LIC. Recent work has examined its variable-rate coding
and bitrate matching mechanisms
\citep{Jia_VRC_2025a,Pan-2025}, robustness under distribution shifts
and adversarial perturbations
\citep{kovalev2024exploring}, adaptation to remote-sensing content
\citep{zhang2025efficient}, and compressed-domain inference
\citep{alkhateeb2024jpeg}. These studies show that standardization does not
terminate codec development; rather, it provides a common framework within
which practical coding functions can be evaluated and extended.

\subsubsection{IEEE 1857.11}
\label{sec:ieee1857}

IEEE 1857.11-2024 provides another standardization path for neural
network-based image coding \citep{IEEE1857}. The standard was developed
through the IEEE 1857.11 Subworking Group and formally published in 2024.
A distinguishing characteristic is its use of multiple profiles targeting
different complexity--performance operating regions.

\textbf{Complexity-oriented profiles.}
IEEE 1857.11 defines Base, Main, and High profiles. The corresponding
reference implementations include BEE, NIC, and iWave
\citep{zhang2023end,chen2021end,ma2019iwave,ma2020end}. These profiles
illustrate different design points in the R--D--complexity space.

The Base profile emphasizes low-complexity implementation and incorporates
efficient mechanisms such as wavefront parallel processing, adaptive
quantization, and feature scaling. The Main profile adopts a more expressive
learned coding architecture with non-local processing, hierarchical priors,
and context-based entropy modeling, providing a stronger balance between
coding efficiency and complexity. The High profile employs a learned
wavelet-like transform and targets higher compression performance. Rather
than enforcing a single neural architecture for every deployment scenario,
the profile structure explicitly recognizes that different applications
have different computational budgets.

\textbf{Common evaluation conditions.}
The verification of IEEE 1857.11 considers images across multiple
resolutions, including ultra-high-resolution content and commonly used
benchmark material. Common training and evaluation conditions are used to
reduce variations caused by training data and testing protocols. Under its
reported evaluation settings, the standard demonstrates substantial
bitrate savings relative to BPG under PSNR and MS-SSIM-based comparison
\citep{IEEE1857}.

This standardized evaluation environment is important because cross-paper
comparison of LIC models can otherwise be strongly affected by training
data, color conversion, resolution handling, entropy-coder implementation,
and anchor configuration. The increasing use of IEEE 1857.11 as a
reference point in subsequent coding studies
\citep{ye2025rate,li2025ustc} further illustrates the value of a common
standardized baseline beyond the individual models contained in the
standard.

\subsubsection{Discussion and Perspective}
\label{sec:standard_discussion}

The emergence of JPEG AI and IEEE 1857.11 marks an important transition
from research-oriented neural compression models to interoperable codec
systems. Unlike cross-paper R--D evaluation, standardization requires
reproducible bitstream syntax, decoder behavior, numerical consistency,
complexity profiles, and conformance procedures. These requirements make
implementation constraints and interoperability first-class codec-design
considerations rather than secondary engineering details.

A remaining challenge is how to reconcile rapid model evolution with
long-term decoding compatibility. Learned codecs can continue to improve
through new architectures, training strategies, and model updates, whereas
a coding standard must preserve stable decoder behavior across deployment
cycles. Future standardization therefore needs mechanisms that accommodate
model and functionality evolution without fragmenting the decoding
ecosystem. Overall, the practical adoption of learned image coding will
depend not only on compression efficiency, but also on reproducibility,
compatibility, and sustainable codec evolution.

\begin{figure}[t]
\centering
\includegraphics[width=0.8\linewidth]{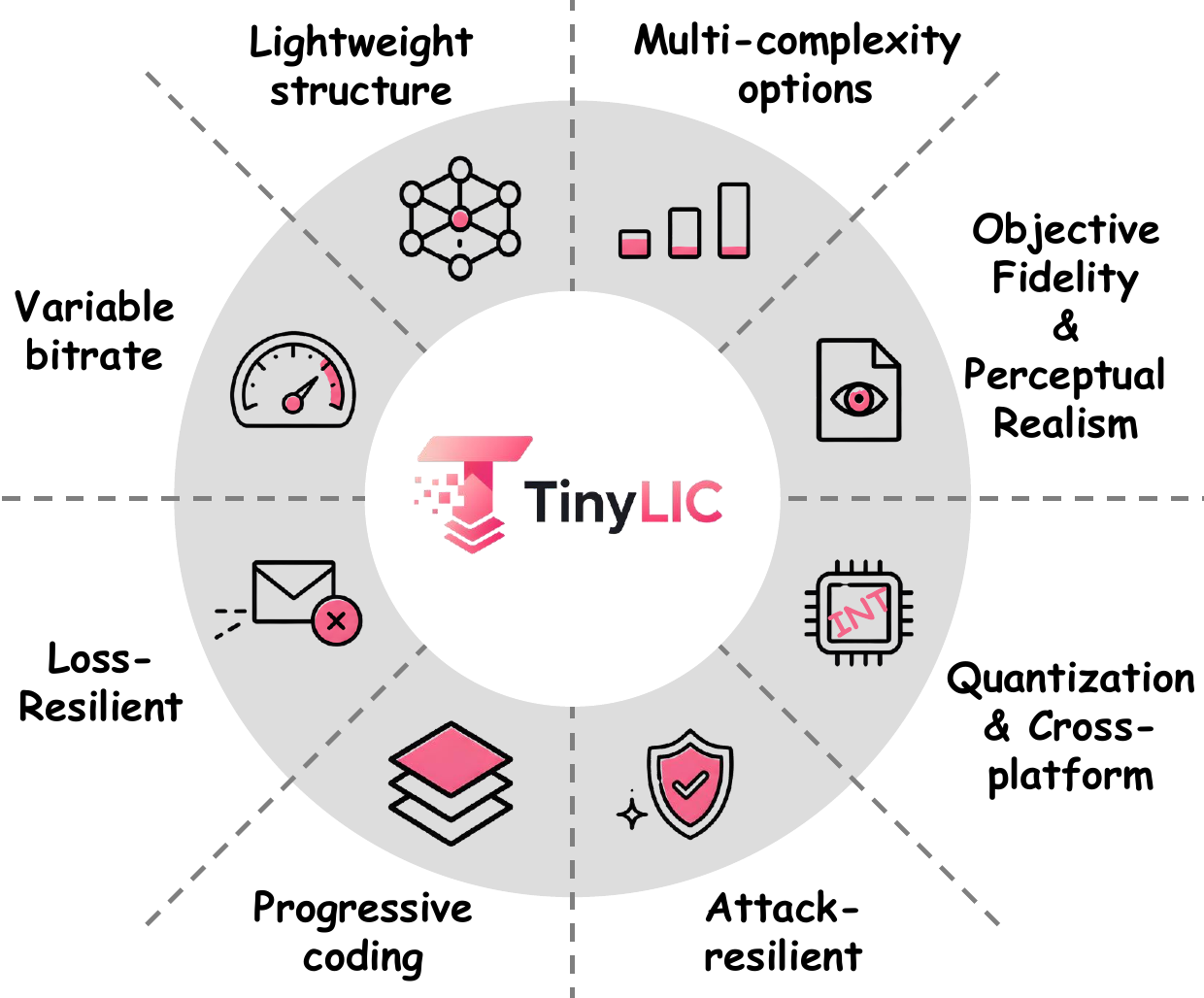}
\caption{TinyLIC as a practical benchmark for learned image coding.}
\label{fig:benchmark}
\end{figure}

% \section{TinyLIC: A Reproducible Practical Benchmark}
% \label{sec:tinylic}
% This section presents \textbf{TinyLIC} as a reproducible case study for deployment-oriented learned image coding. Its role is to operationalize the design questions reviewed above: how R--D performance changes when complexity, rate control, numerical consistency, robustness, progressive delivery, and network resilience are treated as coupled deployment constraints. TinyLIC therefore serves as an open, lightweight reference implementation and benchmark that makes these trade-offs measurable under a common architecture and evaluation protocol. Unlike recent codecs optimized primarily for raw R--D performance, TinyLIC provides a transparent and extensible baseline for analyzing rate, distortion, complexity, controllability, robustness, and deployment behavior under reproducible conditions. The purpose of this section is therefore explanatory: \lm{TinyLIC is not intended as a standalone state-of-the-art codec, but as a reproducible reference platform that operationalizes the practical requirements reviewed in Sec. 3 under a unified architecture.}

\begin{figure*}[!ht]
\centering
\includegraphics[width=0.96\linewidth]{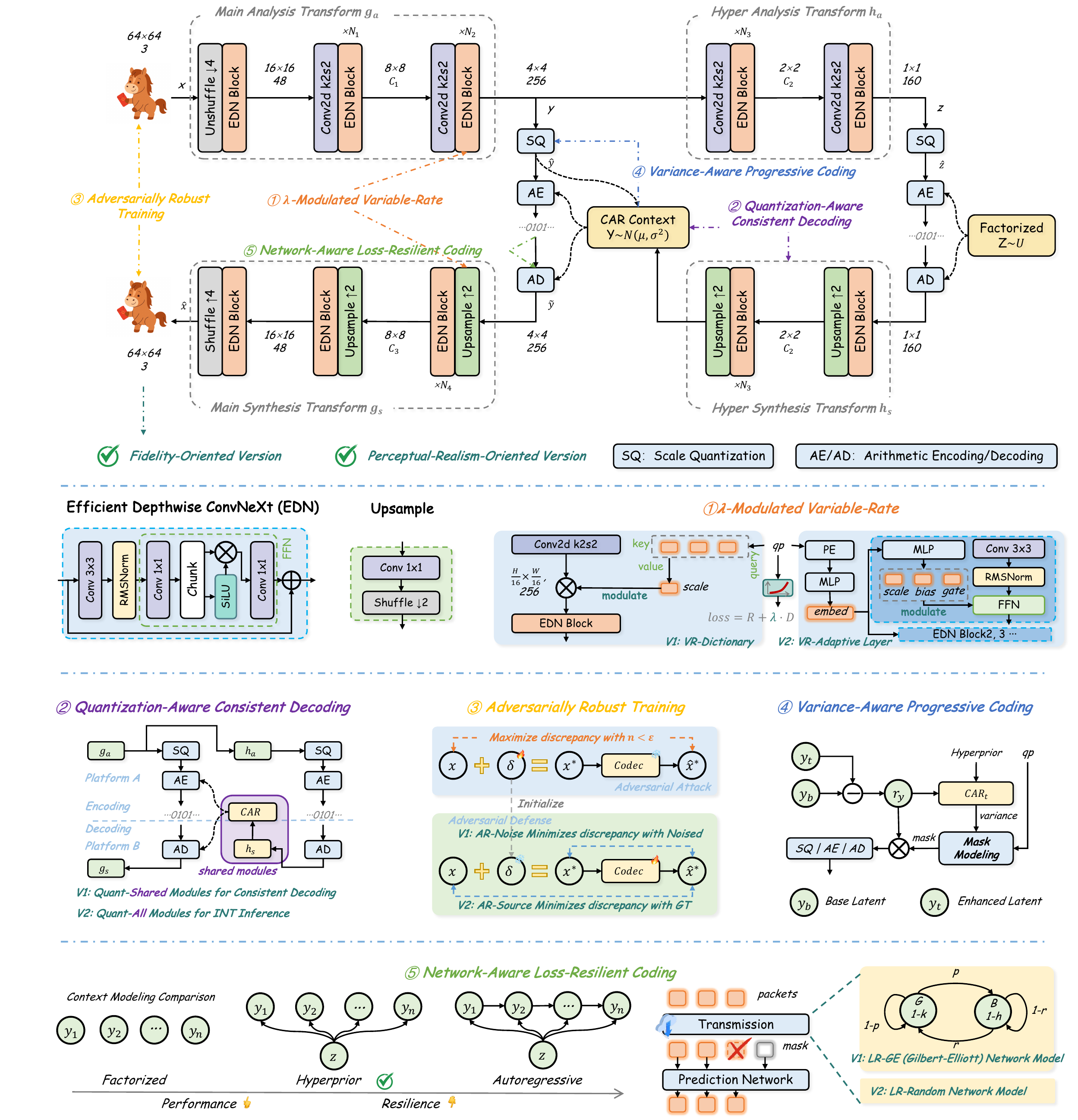}
\caption{TinyLIC employs a lightweight VAE-based architecture using a channel-autoregressive entropy model and efficient depthwise ConvNeXt-style blocks. It further integrates five configurable tools for variable-rate coding, consistent decoding, adversarially robust training, progressive transmission, and loss resilience.}
\label{fig:tinylic_overview}
\end{figure*}

\definecolor{distbg}{RGB}{232,244,238}
\definecolor{disttext}{RGB}{55,120,95}
\definecolor{percbg}{RGB}{240,235,248}
\definecolor{perctext}{RGB}{112,70,155}
\definecolor{groupbg}{RGB}{247,247,247}

\section{TinyLIC: A Reproducible Practical Benchmark}
\label{sec:tinylic}

The practical requirements reviewed in Sec.~\ref{sec:practical} are
typically studied in isolation and often evaluated using different codec
architectures, training settings, and test protocols. This makes it
difficult to determine whether an observed gain or degradation originates
from the deployment mechanism itself or from differences in the underlying
compression model. A common experimental platform is therefore useful for
studying these system-level properties under controlled conditions.

To this end, we introduce \emph{TinyLIC} as an open and reproducible
benchmark for deployment-oriented learned image coding. TinyLIC provides a
lightweight reference architecture together with configurable complexity
levels and modular support for representative practical functions. Its
purpose is to make the interactions among compression efficiency,
computational complexity, controllability, numerical consistency, and
robustness measurable within a shared codec framework.

As illustrated in Fig.~\ref{fig:benchmark}, TinyLIC exposes three
complementary dimensions of practical codec design. First, the backbone can
be scaled across multiple computational budgets, enabling direct analysis of
the R--D--complexity relationship. Second, both fidelity-oriented and
perceptual-realism-oriented optimization are supported, allowing the same
framework to cover different reconstruction objectives. Third, practical
functions corresponding to the deployment issues reviewed in Sec.~\ref{sec:practical} can be
enabled as configurable modules, including variable-rate coding,
quantization-aware consistent decoding, adversarially robust training,
progressive coding, and network-aware loss-resilient coding.

This organization allows each practical function to be evaluated against a
common baseline rather than across independently developed codecs. In
particular, we can examine not only whether a deployment mechanism works,
but also what cost it introduces in bitrate, distortion, computational
complexity, or clean-condition performance. TinyLIC therefore serves as a
controlled experimental platform for studying the trade-offs that arise when
a learned compression model is extended toward a practical codec.

The remainder of this section first introduces the design philosophy and
lightweight architecture of TinyLIC, followed by the implementation of its
deployment-oriented tools. We then evaluate the resulting
rate--distortion--perception performance, practical functionality, and model
complexity under a common experimental protocol.

\subsection{Design Philosophy}
\label{sec:tinylic_design}

TinyLIC is designed around three principles: \emph{simplicity},
\emph{diagnosability}, and \emph{extensibility}.

\textbf{Simplicity.}
The core codec is intentionally kept compact so that its computational
structure and major sources of coding gain remain transparent. Instead of
relying on heavily over-parameterized transforms or complex attention
modules, TinyLIC adopts a lightweight VAE-based backbone with a controlled
entropy-modeling structure. This makes it possible to scale model complexity
over a wide range without changing the fundamental coding pipeline and
provides a common reference for rate--distortion--complexity analysis.

\textbf{Diagnosability.}
Deployment-oriented functions are implemented as configurable modules rather
than being tightly coupled with the backbone. Variable-rate coding,
consistent decoding, adversarial robustness, progressive transmission, and
loss resilience can therefore be activated independently or in combination.
This design makes the cost of individual system requirements easier to
isolate. For example, the R--D degradation introduced by full integer
inference can be distinguished from that caused by resilience-oriented
entropy modeling rather than being hidden by differences between unrelated
codec architectures.

\textbf{Extensibility.}
TinyLIC provides a reusable experimental infrastructure rather than a
closed architecture. Its transform, entropy model, rate-control mechanism,
and deployment modules can be replaced or extended while preserving the
overall evaluation pipeline. New lightweight blocks, probability models,
quantization strategies, robustness objectives, or communication-aware
mechanisms can therefore be studied under the same benchmark setting.

These principles motivate the separation between the \emph{core codec} and
the \emph{practical tools} illustrated in Fig.~\ref{fig:tinylic_overview}.
The core codec defines the representation and entropy-coding pipeline,
whereas the optional tools introduce specific deployment capabilities.
Keeping the underlying architecture fixed whenever possible allows their
R--D and complexity costs to be measured directly.

\subsubsection{Efficient Architecture}
\label{sec:tinylic_architecture}

As illustrated in Fig.~\ref{fig:tinylic_overview}, TinyLIC follows a
lightweight VAE-based coding pipeline.

\textbf{Lightweight transform.}
The main transforms are constructed primarily from convolutional operations
rather than computationally intensive attention modules. The basic unit is
an efficient depthwise ConvNeXt-style block, termed the \emph{EDN block},
which combines lightweight spatial filtering, normalization, pointwise
projection, and a compact feed-forward path. Downsampling and upsampling are
implemented using simple convolutional and shuffle-based operations. The
same basic topology is retained across different complexity configurations,
making complexity scaling easier to interpret.

\textbf{Lightweight entropy modeling.}
TinyLIC models the main latent representation using a Gaussian likelihood
together with a channel-autoregressive (CAR) context model
\citep{minnen2020channel}:
\begin{equation}
    p(\hat{{y}}\mid\hat{{z}})
    =
    \prod_{t=1}^{T}
    p\!\left(
    \hat{{y}}_t
    \mid
    \hat{{y}}_{<t},
    h_s(\hat{{z}})
    \right),
    \qquad
    \hat{{y}}_t
    \sim
    \mathcal{N}(\mu_t,\sigma_t^2),
    \label{eq:tinylic_car}
\end{equation}
where $\hat{{y}}_t$ denotes the $t$-th channel group and $T$ is the
number of autoregressive steps. TinyLIC uses a four-step CAR configuration
as the default operating point.

CAR is adopted as a compromise between probability-modeling capacity and
decoding complexity. In contrast to spatial autoregressive models based on
masked convolutions, channel grouping exposes explicit control over the
number of sequential steps and allows computation associated with unused
channel groups to be avoided. This property is particularly useful when
constructing lightweight codecs with predictable decoding complexity.

\textbf{Complexity scaling.}
TinyLIC provides three configurations---\emph{Nano}, \emph{Small}, and
\emph{Base}---covering approximately 10, 20, and 50 KMACs/pixel of
decoding complexity, respectively. The configurations are obtained by
scaling channel width and block depth while preserving the overall codec
topology. This creates a controlled complexity axis for studying how
compression efficiency changes as additional computational capacity is
introduced.

\textbf{Fidelity- and perceptual-realism-oriented variants.}
The same architecture supports two reconstruction objectives. The
fidelity-oriented version follows conventional R--D optimization using MSE
distortion and is evaluated primarily by PSNR. The
perceptual-realism-oriented version additionally incorporates feature-space
losses following LPIPS \citep{zhang2018unreasonable} and adversarial
optimization to improve visual realism at low bitrates. These two variants
allow the effect of computational complexity to be studied under both
fidelity- and perception-oriented objectives without changing the basic
codec architecture.

\subsubsection{Practical Tools}
\label{sec:tinylic_tools}

Beyond the lightweight backbone, TinyLIC integrates five configurable tools
corresponding to the deployment issues reviewed in Sec.~\ref{sec:practical}:
variable-rate coding, quantization-aware consistent decoding,
adversarially robust training, progressive coding, and network-aware
loss-resilient coding. These modules are optional rather than mandatory,
reflecting the fact that practical functionality may introduce different
R--D or implementation costs.

\textbf{Variable-rate coding.}
TinyLIC adopts $\lambda$-conditioned feature modulation to support multiple
R--D operating points within a single model, following the general
modulation paradigm discussed in Sec.~\ref{sec:variable_rate}
\citep{chen2020variable}. Given an intermediate feature ${f}$,
the modulation is formulated as
\begin{equation}
    {f}'
    =
    {f}
    +
    \gamma(\lambda)
    \odot
    \left[
        \alpha(\lambda)\odot{f}
        +
        \beta(\lambda)
    \right],
    \label{eq:tinylic_vr}
\end{equation}
where $\alpha(\lambda)$, $\beta(\lambda)$, and $\gamma(\lambda)$ denote
learned scale, bias, and gate parameters, respectively. Two implementations are considered.

\begin{itemize}
    \item \textbf{VR-Adaptive Layer.}
    A lightweight multilayer perceptron maps $\lambda$ to the modulation
    parameters dynamically, providing continuous control over the operating
    point.

    \item \textbf{VR-Dictionary.}
    A discrete dictionary of modulation parameters is indexed by a quality
    parameter (QP), with each entry associated with a predefined
    $\lambda$. This design provides a simpler implementation while retaining
    effective variable-rate behavior.
\end{itemize}

\textbf{Quantization-aware consistent decoding.}
TinyLIC supports INT8 quantization with channel-wise weight quantization and layer-wise activation quantization. Two configurations are used to distinguish numerical consistency from full integer deployment.

\begin{itemize}
    \item \textbf{Quant-Shared.}
    Only the modules shared by encoder and decoder for entropy-parameter
    generation are quantized. This configuration targets deterministic
    entropy decoding while retaining most transform computation in
    floating point.

    \item \textbf{Quant-All.}
    The complete codec is quantized for INT8 inference, targeting
    accelerator-oriented execution in addition to entropy-decoding
    consistency.
\end{itemize}

This separation directly reflects the distinction discussed in Sec.~\ref{sec:quantization_consistency}:
the numerical scope required for bitstream interoperability can be smaller
than that required for full low-precision inference.

\textbf{Adversarially robust training.}
To examine codec reliability under input perturbations, TinyLIC incorporates
two adversarially robust training configurations inspired by
compression-specific robustness studies \citep{chen2023toward}. Adversarial
samples are generated under a bounded perturbation constraint, including
CW-style optimization \citep{carlini2017towards}.

\begin{itemize}
    \item \textbf{AR-Source.}
    The reconstruction from the adversarial input is encouraged to remain
    close to the clean source image, explicitly promoting recovery of the
    original content.

    \item \textbf{AR-Noise.}
    The reconstruction from the adversarial input is constrained to remain
    consistent with that obtained from a noise-perturbed reference,
    encouraging local stability around the source distribution.
\end{itemize}

Both strategies modify training rather than inference and therefore do not
introduce additional runtime modules.

\textbf{Variance-aware progressive coding.}
TinyLIC introduces progressiveness through latent-element ordering rather
than an independent multi-stage codec. Following variance-aware progressive
coding \citep{presta2025efficient}, the predicted variance of each latent
element is used as a proxy for its coding importance. More informative
elements are transmitted first, while untransmitted elements are replaced
by their predicted means:
\begin{equation}
    \tilde{y}_i
    =
    m_i\hat{y}_i
    +
    (1-m_i)\mu_i,
    \qquad
    m_i\in\{0,1\},
    \label{eq:tinylic_progressive}
\end{equation}
where $m_i$ indicates whether the $i$-th latent element is available.
Increasing the number of transmitted elements therefore progressively
improves reconstruction quality without changing the basic VAE pipeline.

\textbf{Network-aware loss-resilient coding.}
Finally, TinyLIC considers unreliable transmission in which compressed
representations may experience packet loss. In the loss-resilient
configuration, the channel-autoregressive dependency is removed and a
hyperprior-based entropy model is used instead. This choice reflects the
performance--resilience trade-off discussed in Sec.~\ref{sec:robustness}: autoregressive
context improves compression efficiency under error-free transmission but
can propagate decoding failures when context symbols are missing
\citep{sha2025towards,wang2025resicomp}.

Packet loss is explicitly simulated during training using two channel
models.

\begin{itemize}
    \item \textbf{LR-GE.}
    A Gilbert--Elliott two-state Markov model is used to simulate bursty
    packet loss by alternating between good and bad channel states.

    \item \textbf{LR-Random.}
    Packet losses are sampled independently according to a fixed loss
    probability, providing a simpler stochastic-loss baseline.
\end{itemize}

The five tools intentionally impose different constraints on the same
underlying codec. Variable-rate modulation mainly changes the operating
condition; selective and full quantization modify numerical execution;
adversarial training changes the optimization distribution; progressive
coding constrains latent ordering; and loss-resilient coding alters entropy
dependencies. The benchmark evaluation in Sec.~\ref{sec:tinylic_evaluation} uses these configurations
to quantify how different deployment requirements affect compression
efficiency under a common architectural basis.

\begin{table*}[!htbp]
  \centering
  \scriptsize
  \caption{Performance of TinyLIC under different complexity budgets and practical tool configurations. Fidelity-oriented models are evaluated by PSNR BD-rate using VTM Intra as the H.266/VVC image-profile anchor, while perceptual-realism-oriented models are evaluated by LPIPS and FID BD-rate using MS-ILLM as the anchor. Lower values indicate better performance.}
  \renewcommand{\arraystretch}{1.10}
  \resizebox{\textwidth}{!}{%
  \begin{tabular}{llcccccccccc}
    \toprule
          &       & \multicolumn{4}{c}{\textcolor{disttext}{\textbf{Objective-Fidelity-Oriented}} } & \multicolumn{6}{c}{\textcolor{perctext}{\textbf{Perceptual-Realism-Oriented }}} \\
\cmidrule(lr){3-6}\cmidrule(lr){7-12}
          &       & \multicolumn{4}{c}{\cellcolor{distbg}\textcolor{disttext}{\textbf{PSNR BD-rate (\%)}}} & \multicolumn{4}{c}{\cellcolor{percbg}\textcolor{perctext}{\textbf{LPIPS BD-rate (\%)}}} & \multicolumn{2}{c}{\cellcolor{percbg}\textcolor{perctext}{\textbf{FID BD-rate (\%)}}} \\
\cmidrule(lr){3-6}\cmidrule(lr){7-10}\cmidrule(lr){11-12}
    Methods & KMACs/pixel & \cellcolor{distbg}\rotatebox{45}{\textcolor{disttext}{\textbf{Kodak}}} & \cellcolor{distbg}\rotatebox{45}{\textcolor{disttext}{\textbf{CLIC}}} & \cellcolor{distbg}\rotatebox{45}{\textcolor{disttext}{\textbf{Tecnick}}} & \cellcolor{distbg}\rotatebox{45}{\textcolor{disttext}{\textbf{JPEG AI}}} & \cellcolor{percbg}\rotatebox{45}{\textcolor{perctext}{\textbf{Kodak}}} & \cellcolor{percbg}\rotatebox{45}{\textcolor{perctext}{\textbf{CLIC}}} & \cellcolor{percbg}\rotatebox{45}{\textcolor{perctext}{\textbf{Tecnick}}} & \cellcolor{percbg}\rotatebox{45}{\textcolor{perctext}{\textbf{JPEG AI}}} & \cellcolor{percbg}\rotatebox{45}{\textcolor{perctext}{\textbf{Tecnick}}} & \cellcolor{percbg}\rotatebox{45}{\textcolor{perctext}{\textbf{JPEG AI}}} \\
    \midrule
    \rowcolor{groupbg}
    \multicolumn{12}{l}{\textcolor[rgb]{ .502,  .502,  .502}{\textit{\textbf{Anchor}}}} \\
    VTM Intra 22.0 (H.266/VVC image-profile anchor)~\citep{bross2021overview} &       & \cellcolor{distbg}0.00 & \cellcolor{distbg}0.00 & \cellcolor{distbg}0.00 & \cellcolor{distbg}0.00 & \cellcolor{percbg}/ & \cellcolor{percbg}/ & \cellcolor{percbg}/ & \cellcolor{percbg}/ & \cellcolor{percbg}/ & \cellcolor{percbg}/ \\
    BPG (H.265/HEVC image-profile anchor)~\citep{bellard2016bpg} &       & \cellcolor{distbg}30.10 & \cellcolor{distbg}41.87 & \cellcolor{distbg}33.94 & \cellcolor{distbg}30.65 & \cellcolor{percbg}/ & \cellcolor{percbg}/ & \cellcolor{percbg}/ & \cellcolor{percbg}/ & \cellcolor{percbg}/ & \cellcolor{percbg}/ \\
    ELIC~\citep{he2022elic} & 573      & \cellcolor{distbg}-3.22 & \cellcolor{distbg}-3.89 & \cellcolor{distbg}-4.57 & \cellcolor{distbg}/ & \cellcolor{percbg}/ & \cellcolor{percbg}/ & \cellcolor{percbg}/ & \cellcolor{percbg}/ & \cellcolor{percbg}/ & \cellcolor{percbg}/ \\
    EVC-small~\citep{wang2023evc} & 52      & \cellcolor{distbg}6.9 & \cellcolor{distbg}/ & \cellcolor{distbg}/ & \cellcolor{distbg}/ & \cellcolor{percbg}/ & \cellcolor{percbg}/ & \cellcolor{percbg}/ & \cellcolor{percbg}/ & \cellcolor{percbg}/ & \cellcolor{percbg}/ \\
    AsymLLIC~\citep{wang2024asymllic} & 130      & \cellcolor{distbg}0.65 & \cellcolor{distbg}/ & \cellcolor{distbg}/ & \cellcolor{distbg}/ & \cellcolor{percbg}/ & \cellcolor{percbg}/ & \cellcolor{percbg}/ & \cellcolor{percbg}/ & \cellcolor{percbg}/ & \cellcolor{percbg}/ \\
    ShiftLIC-Small~\citep{bao2025shiftlic} & 114      & \cellcolor{distbg}15.56 & \cellcolor{distbg}11.18 & \cellcolor{distbg}10.67 & \cellcolor{distbg}/ & \cellcolor{percbg}/ & \cellcolor{percbg}/ & \cellcolor{percbg}/ & \cellcolor{percbg}/ & \cellcolor{percbg}/ & \cellcolor{percbg}/ \\
    CSLIC-3~\citep{wang2026distilling} & 125      & \cellcolor{distbg}-0.94 & \cellcolor{distbg}3.48 & \cellcolor{distbg}2.75 & \cellcolor{distbg}/ & \cellcolor{percbg}/ & \cellcolor{percbg}/ & \cellcolor{percbg}/ & \cellcolor{percbg}/ & \cellcolor{percbg}/ & \cellcolor{percbg}/ \\
    CSLIC-1~\citep{wang2026distilling} & 77      & \cellcolor{distbg}5.37 & \cellcolor{distbg}10.18 & \cellcolor{distbg}10.12 & \cellcolor{distbg}/ & \cellcolor{percbg}/ & \cellcolor{percbg}/ & \cellcolor{percbg}/ & \cellcolor{percbg}/ & \cellcolor{percbg}/ & \cellcolor{percbg}/ \\
    ShallowNTC~\citep{yang2023computationally} & 20      & \cellcolor{distbg}22.22 & \cellcolor{distbg}30.36 & \cellcolor{distbg}/ & \cellcolor{distbg}/ & \cellcolor{percbg}/ & \cellcolor{percbg}/ & \cellcolor{percbg}/ & \cellcolor{percbg}/ & \cellcolor{percbg}/ & \cellcolor{percbg}/ \\
    FastNIC~\citep{zhang2025learning} & 10      & \cellcolor{distbg}26.00 & \cellcolor{distbg}34.26 & \cellcolor{distbg}35.61 & \cellcolor{distbg}/ & \cellcolor{percbg}/ & \cellcolor{percbg}/ & \cellcolor{percbg}/ & \cellcolor{percbg}/ & \cellcolor{percbg}/ & \cellcolor{percbg}/ \\
    MS-ILLM~\citep{muckley2023improving} & 600      & \cellcolor{distbg}/ & \cellcolor{distbg}/ & \cellcolor{distbg}/ & \cellcolor{distbg}/ & \cellcolor{percbg}0.00 & \cellcolor{percbg}0.00 & \cellcolor{percbg}0.00 & \cellcolor{percbg}0.00 & \cellcolor{percbg}0.00 & \cellcolor{percbg}0.00 \\
    HiFiC~\citep{mentzer2020high} &600       & \cellcolor{distbg}/ & \cellcolor{distbg}/ & \cellcolor{distbg}/ & \cellcolor{distbg}/ & \cellcolor{percbg}50.30 & \cellcolor{percbg}71.01 & \cellcolor{percbg}50.22 & \cellcolor{percbg}66.97 & \cellcolor{percbg}74.05 & \cellcolor{percbg}68.73 \\
    \midrule
    \rowcolor{groupbg}
    \multicolumn{12}{l}{\textcolor[rgb]{ .502,  .502,  .502}{\textit{\textbf{Fixed-Rate TinyLIC}}}} \\
    Nano  & 10   & \cellcolor{distbg}14.78 & \cellcolor{distbg}14.68 & \cellcolor{distbg}12.76 & \cellcolor{distbg}13.86 & \cellcolor{percbg}30.92 & \cellcolor{percbg}97.47 & \cellcolor{percbg}83.49 & \cellcolor{percbg}87.98 & \cellcolor{percbg}91.57 & \cellcolor{percbg}40.62 \\
    Small & 20   & \cellcolor{distbg}5.43 & \cellcolor{distbg}4.63 & \cellcolor{distbg}2.73 & \cellcolor{distbg}6.33 & \cellcolor{percbg}20.96 & \cellcolor{percbg}78.66 & \cellcolor{percbg}60.03 & \cellcolor{percbg}63.81 & \cellcolor{percbg}76.21 & \cellcolor{percbg}41.86 \\
    Base  & 50   & \cellcolor{distbg}-0.79 & \cellcolor{distbg}-1.76 & \cellcolor{distbg}-4.21 & \cellcolor{distbg}0.91 & \cellcolor{percbg}15.22 & \cellcolor{percbg}51.09 & \cellcolor{percbg}38.85 & \cellcolor{percbg}44.95 & \cellcolor{percbg}35.47 & \cellcolor{percbg}30.72 \\
    \midrule
    \rowcolor{groupbg}
    \multicolumn{12}{l}{\textcolor[rgb]{ .502,  .502,  .502}{\textit{\textbf{Variable-Rate TinyLIC}}}} \\
    VR-Adaptive Layer & 50   & \cellcolor{distbg}-1.50 & \cellcolor{distbg}-2.75 & \cellcolor{distbg}-5.02 & \cellcolor{distbg}0.29 & \cellcolor{percbg}14.75 & \cellcolor{percbg}49.17 & \cellcolor{percbg}41.30 & \cellcolor{percbg}40.02 & \cellcolor{percbg}28.35 & \cellcolor{percbg}26.47 \\
    VR-Dictionary & 50   & \cellcolor{distbg}-0.94 & \cellcolor{distbg}-2.21 & \cellcolor{distbg}-4.26 & \cellcolor{distbg}0.96 & \cellcolor{percbg}15.20 & \cellcolor{percbg}49.80 & \cellcolor{percbg}41.57 & \cellcolor{percbg}40.66 & \cellcolor{percbg}28.97 & \cellcolor{percbg}28.75 \\
    \midrule
    \rowcolor{groupbg}
    \multicolumn{12}{l}{\textcolor[rgb]{ .502,  .502,  .502}{\textit{\textbf{Variable-Rate (VR-Dictionary) + Consistent Decoding}}}} \\
    Quant-All (INT8) & 50   & \cellcolor{distbg}2.21 & \cellcolor{distbg}2.49 & \cellcolor{distbg}-0.22 & \cellcolor{distbg}4.74 & \cellcolor{percbg}17.16 & \cellcolor{percbg}60.02 & \cellcolor{percbg}43.20 & \cellcolor{percbg}51.52 & \cellcolor{percbg}32.59 & \cellcolor{percbg}28.60 \\
    Quant-Shared (INT8) & 50   & \cellcolor{distbg}-0.41 & \cellcolor{distbg}-1.45 & \cellcolor{distbg}-3.68 & \cellcolor{distbg}1.52 & \cellcolor{percbg}13.98 & \cellcolor{percbg}50.06 & \cellcolor{percbg}38.38 & \cellcolor{percbg}45.32 & \cellcolor{percbg}26.80 & \cellcolor{percbg}29.22 \\
    \midrule
    \rowcolor{groupbg}
    \multicolumn{12}{l}{\textcolor[rgb]{ .502,  .502,  .502}{\textit{\textbf{Variable-Rate (VR-Dictionary) + Consistent Decoding (Quant-Shared) + Progressive Coding / Attack-Resilient /  Loss-Resilient}}}} \\
    PC-Variance Aware & 50   & \cellcolor{distbg}3.15 & \cellcolor{distbg}-1.04 & \cellcolor{distbg}-1.71 & \cellcolor{distbg}2.55 & \cellcolor{percbg}21.88 & \cellcolor{percbg}67.02 & \cellcolor{percbg}52.24 & \cellcolor{percbg}55.09 & \cellcolor{percbg}44.01 & \cellcolor{percbg}44.08 \\
    AR-Source & 50   & \cellcolor{distbg}0.75 & \cellcolor{distbg}-0.84 & \cellcolor{distbg}-2.69 & \cellcolor{distbg}2.67 & \cellcolor{percbg}15.45 & \cellcolor{percbg}61.89 & \cellcolor{percbg}42.41 & \cellcolor{percbg}51.76 & \cellcolor{percbg}37.63 & \cellcolor{percbg}34.43 \\
    AR-Noise & 50   & \cellcolor{distbg}-0.20 & \cellcolor{distbg}-1.36 & \cellcolor{distbg}-3.36 & \cellcolor{distbg}1.80 & \cellcolor{percbg}14.55 & \cellcolor{percbg}60.05 & \cellcolor{percbg}41.07 & \cellcolor{percbg}49.69 & \cellcolor{percbg}35.19 & \cellcolor{percbg}32.98 \\
    LR-GE  (20\%) & 50   & \cellcolor{distbg}9.78 & \cellcolor{distbg}8.52 & \cellcolor{distbg}6.22 & \cellcolor{distbg}8.73 & \cellcolor{percbg}33.85 & \cellcolor{percbg}81.91 & \cellcolor{percbg}72.88 & \cellcolor{percbg}58.90 & \cellcolor{percbg}56.35 & \cellcolor{percbg}41.32 \\
    LR-Random (20\%) & 50   & \cellcolor{distbg}9.69 & \cellcolor{distbg}8.54 & \cellcolor{distbg}6.01 & \cellcolor{distbg}8.53 & \cellcolor{percbg}33.52 & \cellcolor{percbg}81.01 & \cellcolor{percbg}71.41 & \cellcolor{percbg}57.30 & \cellcolor{percbg}54.54 & \cellcolor{percbg}39.53 \\
    \bottomrule
  \end{tabular}}%
  \label{tab:tinylic_performance}
\end{table*}

\begin{figure*}[ht]
\centering
\includegraphics[width=0.96\linewidth]{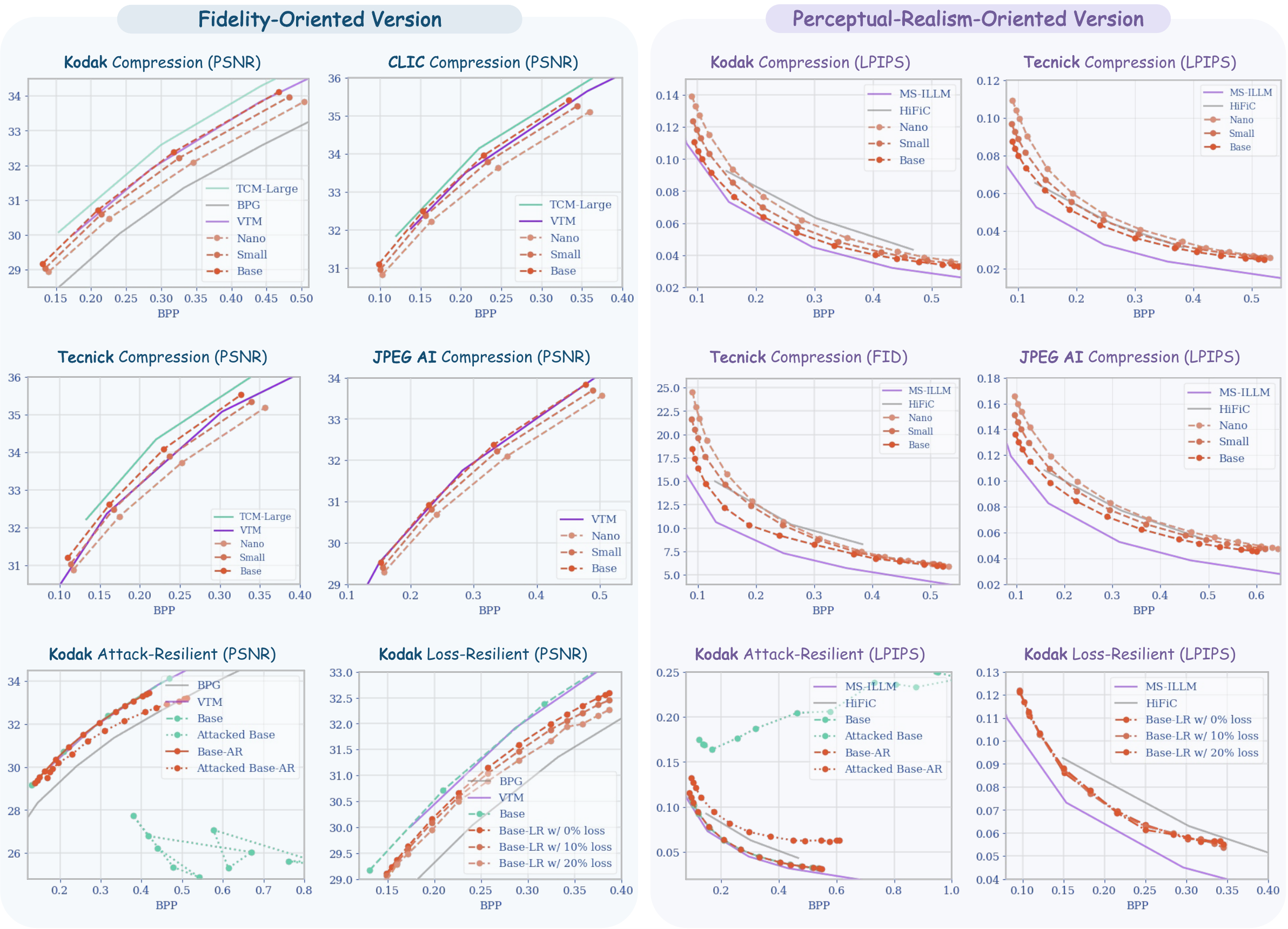}
\caption{R--D and perceptual performance of TinyLIC under different complexity budgets and practical tool configurations.}
\label{fig:tinylic_results}
\end{figure*}

\subsection{Benchmark Evaluation}
\label{sec:tinylic_evaluation}

We next evaluate TinyLIC as a common benchmark for studying the
R--D--complexity trade-offs and deployment-oriented functionalities
introduced above. The evaluation covers both fidelity- and
perceptual-realism-oriented compression, multiple computational budgets,
and the practical tools for variable-rate coding, consistent decoding,
progressive coding, adversarial robustness, and loss resilience. Unless
otherwise specified, these variants share the same underlying TinyLIC
architecture, allowing the effect of individual deployment mechanisms to be
examined under controlled conditions.

\subsubsection{Experimental Setup}
\label{sec:tinylic_setup}

\textbf{Training protocol.}
TinyLIC is trained on the MLIC-100K dataset
\citep{jiang2025mlic++}. Unless otherwise specified, all models are trained
for 400 epochs with a batch size of 32. The initial learning rate is set to
$2\times10^{-4}$ and reduced to $2\times10^{-5}$ after 300 epochs.
Fidelity-oriented models are optimized using the conventional R--D objective
with MSE as the distortion term. The
perceptual-realism-oriented variants additionally incorporate feature-space
and adversarial objectives, as described in Sec.~\ref{sec:tinylic_architecture}.

\textbf{Test datasets.}
We evaluate TinyLIC on Kodak, Tecnick
\citep{asuni2014testimages}, CLIC Professional validation, and the JPEG AI
test set \citep{JPEG-AI}. These datasets span different image resolutions
and content characteristics. Kodak contains 24 images at $768\times512$
resolution, Tecnick contains high-quality $1200\times1200$ images, CLIC
Professional validation contains 41 images with approximately 2K
resolution, and the JPEG AI test set contains 50 natural images ranging
from approximately 1K to 4K resolution.

For high-resolution JPEG AI images, TinyLIC performs inference using
non-overlapping tiles no larger than $1024\times1024$ to satisfy memory
constraints. This should be noted when interpreting comparisons with
conventional codecs that process the complete image at full resolution.

\textbf{Compared methods.}
For fidelity-oriented evaluation, we use VTM Intra 22.0 under the
H.266/VVC image-profile configuration
\citep{bross2021developments} and BPG, corresponding to an
H.265/HEVC image-profile anchor \citep{bellard2016bpg}, as conventional
references. Representative learned codecs include ELIC
\citep{he2022elic} and recent lightweight approaches such as EVC
\citep{wang2023evc}, AsymLLIC \citep{wang2024asymllic}, ShiftLIC
\citep{bao2025shiftlic}, CSLIC \citep{wang2026distilling}, ShallowNTC
\citep{yang2023computationally}, and FastNIC
\citep{zhang2025learning}.

For perceptual-realism-oriented compression, we compare against HiFiC
\citep{mentzer2020high} and MS-ILLM
\citep{muckley2023improving}. HiFiC represents a classical GAN-based neural
codec optimized for perceptual reconstruction, while MS-ILLM serves as the
reference method for the reported perceptual BD-rate comparisons.

For fidelity-oriented evaluation, compression efficiency is reported using
PSNR BD-rate with VTM Intra as the anchor. Perceptual-realism-oriented
performance is evaluated using LPIPS and FID BD-rate with MS-ILLM as the
anchor. Computational complexity is reported in KMACs/pixel.

\subsubsection{Rate--Distortion--Perception Performance}
\label{sec:tinylic_rdp}

Figure~\ref{fig:tinylic_results} and
Table~\ref{tab:tinylic_performance} summarize the performance of TinyLIC
under different computational budgets. The results exhibit a clear
performance--complexity scaling trend.

\textbf{Fidelity-oriented performance.}
Increasing the decoding complexity from approximately 10 to
20 KMACs/pixel substantially improves compression efficiency. On Kodak,
the PSNR BD-rate relative to VTM Intra decreases from $14.78\%$ for
TinyLIC-Nano to $5.43\%$ for TinyLIC-Small. Increasing the complexity to
approximately 50 KMACs/pixel further brings TinyLIC-Base close to or beyond
the VTM Intra anchor, with BD-rates of $-0.79\%$, $-1.76\%$, and
$-4.21\%$ on Kodak, CLIC, and Tecnick, respectively. On the JPEG AI test
set, TinyLIC-Base achieves a BD-rate of $0.91\%$.

The 10-KMACs/pixel Nano model also provides a useful low-complexity
operating point. On Kodak, its $14.78\%$ BD-rate relative to VTM Intra is
substantially lower than the $30.10\%$ of BPG, demonstrating that a compact
learned codec can retain meaningful compression gains under a stringent
computational budget.

For the JPEG AI test set, TinyLIC uses tile-based coding because of memory
constraints, whereas VTM Intra processes the complete image. This protocol
difference places TinyLIC at a disadvantage in the direct R--D comparison
and should therefore be considered when interpreting the corresponding
numbers.

Overall, the results indicate that a compact convolutional transform
combined with lightweight entropy modeling can provide a competitive
R--D--complexity trade-off without relying on computationally intensive
attention modules. At the same time, the incremental gain becomes smaller
from the Small to Base configuration, suggesting diminishing returns as
model capacity increases. A similar tendency has also been observed in
recent scaling studies of substantially larger learned compression models
\citep{li2025scaling}.

\textbf{Perceptual-realism-oriented performance.}
The same TinyLIC backbone can also be optimized for perceptual realism.
Under the reported LPIPS and FID BD-rate protocol, the perceptual variants
consistently improve as the computational budget increases. For example,
the Tecnick FID BD-rate relative to MS-ILLM decreases from $91.57\%$ for
TinyLIC-Nano to $76.21\%$ for TinyLIC-Small and $35.47\%$ for
TinyLIC-Base.

TinyLIC also improves over HiFiC under the reported perceptual metrics,
while a gap remains to the stronger MS-ILLM anchor. These results are not
intended to establish a new perceptual-compression frontier; rather, they
show that the same lightweight architecture can support both fidelity- and
realism-oriented optimization, making it possible to study
rate--distortion--perception behavior under controlled complexity budgets.

\subsubsection{Effect of Practical Tools}
\label{sec:tinylic_tools_eval}

The lower part of Table~\ref{tab:tinylic_performance} evaluates the
deployment-oriented tools introduced in Sec.~\ref{sec:tinylic_tools}.
Because these variants are constructed on a common backbone, their results
make the cost of additional practical functionality directly observable.

\textbf{Variable-rate coding.}
Both variable-rate configurations retain performance close to the
corresponding fixed-rate model. The adaptive-layer variant achieves
PSNR BD-rates of $-1.50\%$, $-2.75\%$, and $-5.02\%$ relative to VTM Intra
on Kodak, CLIC, and Tecnick, respectively, while the dictionary-based
variant obtains $-0.94\%$, $-2.21\%$, and $-4.26\%$.

The apparent improvement over the fixed-rate Base model should be interpreted
carefully. BD-rate is obtained by interpolating a finite set of R--D
operating points. The fixed-rate evaluation contains four operating points,
whereas the variable-rate models are evaluated using 15 points, resulting
in denser coverage of the R--D curve and potentially reducing interpolation
error. The result therefore primarily demonstrates that rate flexibility
can be introduced without a clear R--D penalty.

Although the adaptive-layer variant provides slightly better numerical
performance, the dictionary-based variant offers a simpler implementation
and is used as the basis for the subsequent practical configurations.

\begin{figure*}[ht]
\centering
\includegraphics[width=\linewidth]{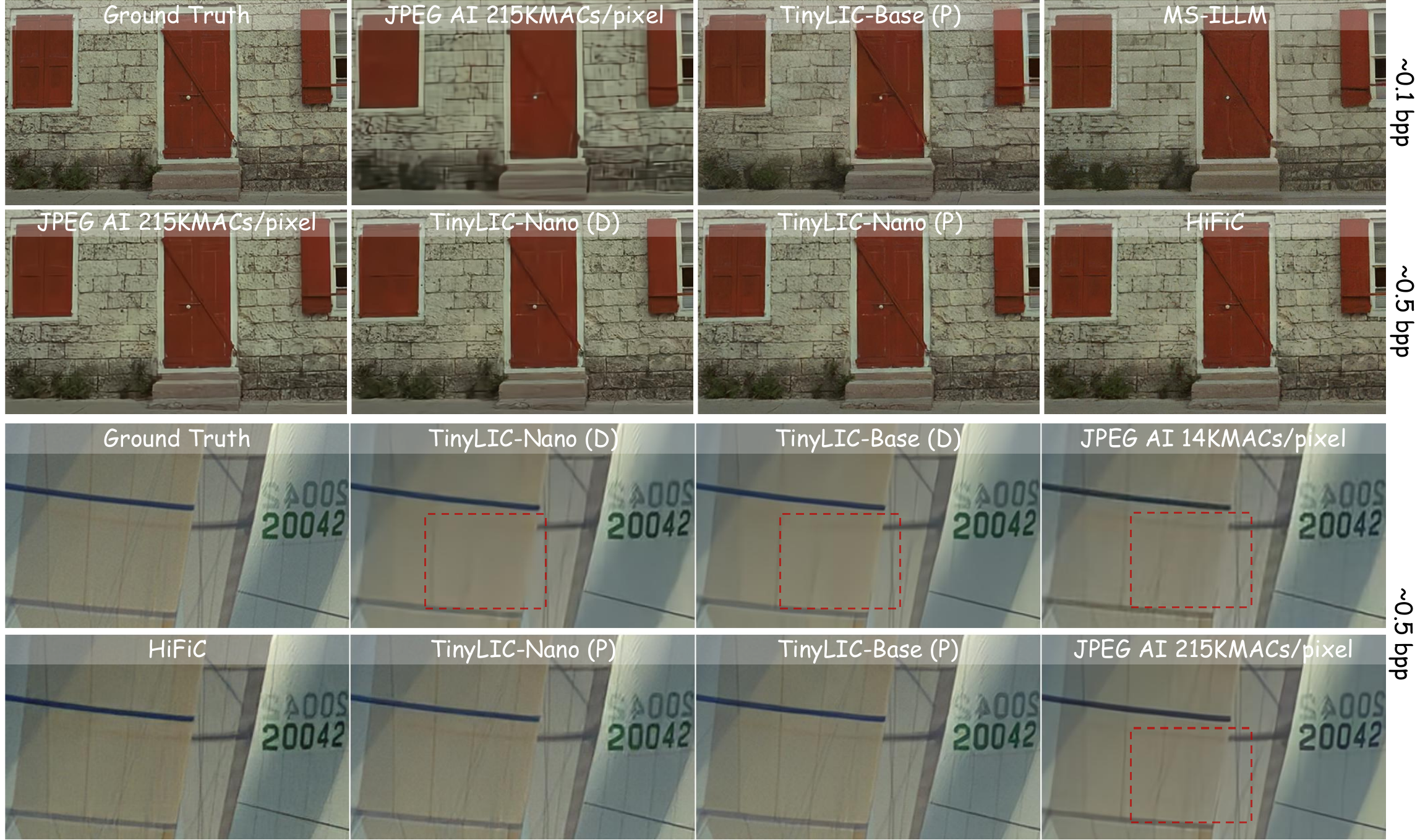}
\caption{Qualitative comparison of TinyLIC with representative codecs. Fidelity-oriented variants (D) preserve structural fidelity, while perceptual-realism-oriented variants (P) recover sharper textures and more visually realistic details.}
\label{fig:tinylic_visual}
\end{figure*}

\textbf{Quantization-aware consistent decoding.}
Quant-Shared quantizes only the modules required for consistent entropy
decoding and introduces a relatively small degradation relative to the
variable-rate baseline. Its PSNR BD-rate on Kodak changes from $-0.94\%$
for VR-Dictionary to $-0.41\%$, while maintaining consistent decoding
across platforms.

Quant-All applies INT8 quantization more broadly to support
hardware-oriented integer inference. This produces a larger R--D penalty,
with a Kodak BD-rate of $2.21\%$. The difference between Quant-Shared and
Quant-All illustrates the distinction discussed in
Sec.~\ref{sec:quantization_consistency}: selective quantization can provide
codec consistency with limited performance loss, whereas full integer
inference introduces a stronger numerical constraint in exchange for
broader hardware efficiency.

\textbf{Progressive coding.}
Variance-aware progressive coding introduces an additional R--D cost because
the latent representation must support an importance-aware transmission
order. The resulting Kodak PSNR BD-rate is $3.15\%$. This degradation
reflects the cost of constraining a representation optimized for complete
decoding to also support meaningful reconstruction from partial
bitstreams.

Importantly, this functionality is introduced without redesigning the
overall codec backbone, allowing the cost of progressiveness to be measured
relative to the same underlying representation.

\textbf{Adversarial robustness.}
Attack-resilient training introduces a smaller clean-image penalty than the
progressive and loss-resilient configurations. AR-Source achieves a Kodak
PSNR BD-rate of $0.75\%$, while AR-Noise reaches $-0.20\%$. AR-Noise also
consistently preserves better clean-image R--D performance than AR-Source
across the reported datasets.

These results illustrate an important property of robustness-oriented
training: improved behavior under perturbations need not require additional
inference-time computation, but it may alter the clean-data optimum learned
by the codec.

\textbf{Loss resilience.}
The loss-resilient variants introduce the largest clean-channel R--D
degradation among the evaluated practical tools. At a training packet-loss
rate of $20\%$, LR-GE and LR-Random achieve Kodak PSNR BD-rates of
$9.78\%$ and $9.69\%$, respectively.

This degradation is largely attributable to the removal of the
autoregressive entropy model in the loss-resilient configurations. While
autoregressive dependencies improve probability estimation under reliable
transmission, they can amplify error propagation when transmitted symbols
are missing. Replacing them with a more independent representation therefore
trades nominal coding efficiency for transmission robustness. The similar
clean-channel behavior of LR-GE and LR-Random further suggests that much
of this cost originates from the resilience-oriented entropy-model
simplification rather than from the particular packet-loss model used
during training.

Taken together, these results illustrate a recurring characteristic of
deployment-oriented LIC: additional functionality is generally associated
with a measurable cost, but the magnitude and source of that cost differ
substantially across requirements. Variable-rate coding and selective
quantization can be introduced with limited R--D degradation, whereas
progressive and especially loss-resilient coding impose stronger structural
constraints on the learned representation.

\subsubsection{Visual Comparison}
\label{sec:tinylic_visual}

Figure~\ref{fig:tinylic_visual} provides qualitative comparisons among
TinyLIC configurations and representative codecs at different operating
points. The examples include fidelity-oriented and
perceptual-realism-oriented TinyLIC variants as well as JPEG AI, HiFiC, and
MS-ILLM.

The fidelity-oriented variants primarily preserve structural and
pixel-level information, whereas perceptual-realism-oriented TinyLIC
produces sharper textures at low bitrates. As expected from perceptual
optimization, these reconstructions may also introduce synthesized
high-frequency details that are not explicitly preserved by the
fidelity-oriented models.

Increasing model complexity from Nano to Base improves both structural
fidelity and visual detail. The comparison further illustrates the intended
role of TinyLIC: rather than relying on a separate architecture for each
optimization target, the same lightweight codec family can expose the
visual consequences of changing the computational budget and reconstruction
objective.

\begin{table}[t]
  \centering
  \footnotesize
  \caption{Complexity and parameter distribution of TinyLIC variants.}
  \label{tab:tinylic_complexity}
  \setlength{\tabcolsep}{1.8pt}
  \renewcommand{\arraystretch}{1.1}
  \begin{tabular}{@{}lccrrrrrrr@{}}
    \toprule
    \multirow{2}{*}{\textbf{TinyLIC}} &
    \multicolumn{2}{c}{\textbf{Complexity}} &
    \multicolumn{7}{c}{\textbf{Parameters (M)}} \\
    \cmidrule(lr){2-3}\cmidrule(lr){4-10}
    &
    \textbf{Enc.} &
    \textbf{Dec.} &
    $\boldsymbol{g_a}$ &
    $\boldsymbol{g_s}$ &
    $\boldsymbol{h_a}$ &
    $\boldsymbol{h_s}$ &
    \textbf{CAR} &
    \textbf{LRP} &
    \textbf{Sum} \\
    \midrule
    Nano  & 11.92 & 9.99  & 0.81 & 0.53 & 0.63 & 0.63 & 0.42 & 0.15 & 3.16  \\
    Small & 20.81 & 19.99 & 1.26 & 1.07 & 0.65 & 0.65 & 0.97 & 0.15 & 4.76  \\
    Base  & 53.00 & 50.76 & 3.45 & 3.05 & 1.38 & 1.38 & 1.17 & 0.15 & 10.57 \\
    \bottomrule
  \end{tabular}
  \begin{tablenotes}[flushleft]
  \item \scriptsize \textit{Note:} Complexity is measured in KMACs/pixel.
  Enc. and Dec. denote encoding and decoding complexity.
  $g_a$/$g_s$: main analysis/synthesis transforms;
  $h_a$/$h_s$: hyper analysis/synthesis transforms;
  CAR: channel autoregressive context model;
  LRP: latent residual prediction~\citep{minnen2020channel}.
  \end{tablenotes}
\end{table}

\begin{figure*}[ht]
\centering
\includegraphics[width=\linewidth]{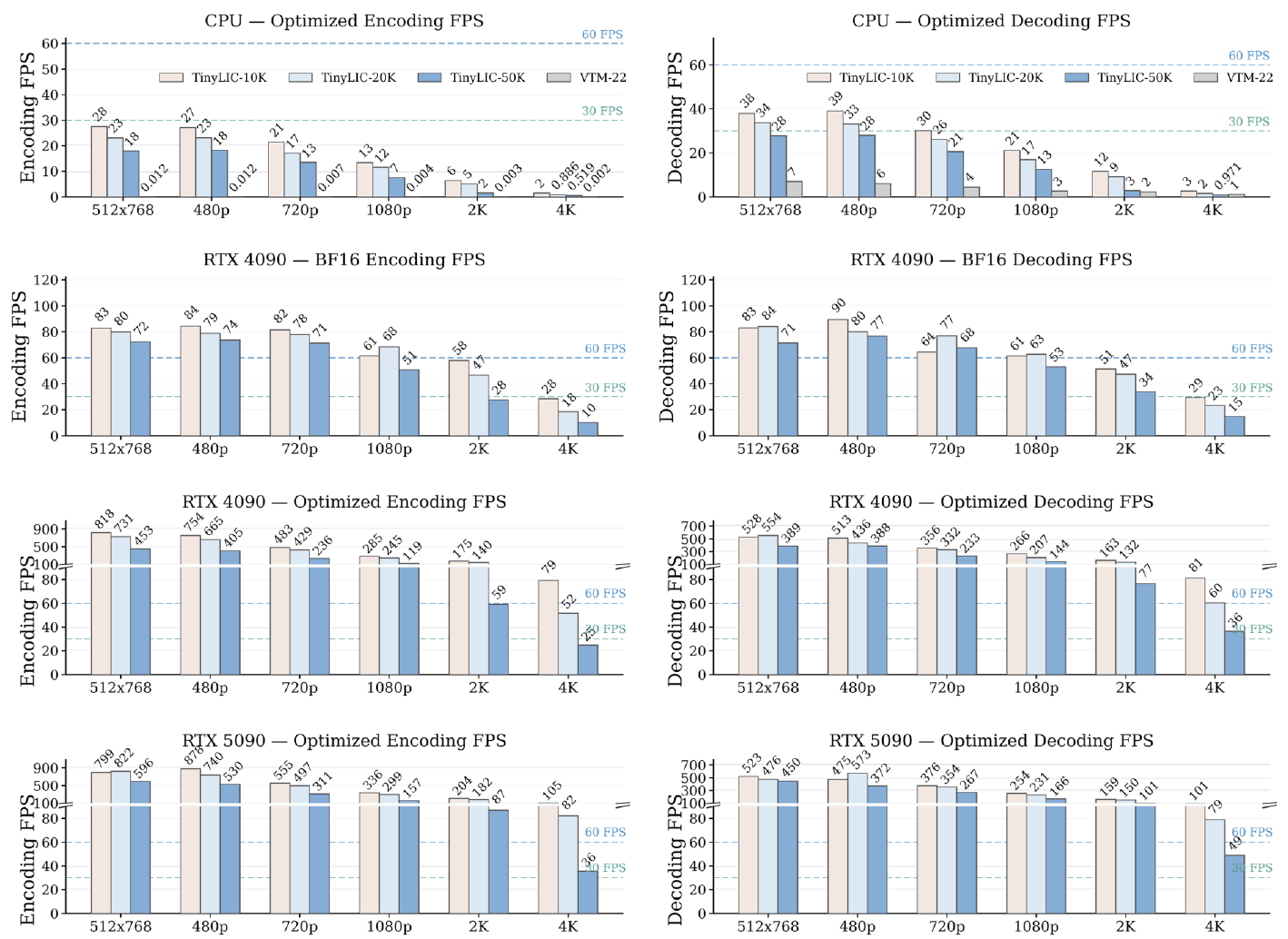}
\caption{Encoding and decoding throughput across resolutions and model scales. The optimized implementation differs from the standard BF16 path by incorporating hardware-aware acceleration, including kernel fusion and specialized high-performance kernels (e.g., CUTLASS on GPU and AMX-assisted execution on CPU), thereby reducing operator-launch and memory-access overheads. Results are reported for TinyLIC-10K (Nano), TinyLIC-20K (Small), and TinyLIC-50K (Base) from $512\times768$ to 4K. Values above the bars indicate FPS, while the dashed lines mark representative 30-FPS and 60-FPS real-time operating points. Broken $y$-axes are adopted where needed for clearer visualization of the wide throughput range.}
\label{fig:fps}
\end{figure*}

\subsubsection{Model Complexity}
\label{sec:tinylic_complexity}

Table~\ref{tab:tinylic_complexity} summarizes the computational complexity and parameter distribution of TinyLIC. Nano, Small, and Base provide a controlled scaling path from approximately 10 to 50 KMACs/pixel decoding complexity while preserving the same codec topology. This enables the R--D improvements in Sec.~\ref{sec:tinylic_rdp} to be attributed directly to increasing model capacity rather than architectural differences.

Beyond theoretical complexity, we further evaluate the practical runtime throughput in Fig.~\ref{fig:fps}. Our optimized implementation incorporates kernel fusion and hardware-specific high-performance kernels, including CUTLASS-based acceleration on GPUs~\citep{cutlass} and AMX-assisted execution on CPUs~\citep{intel_amx}, and therefore better reflects the practical deployment potential of TinyLIC than the standard BF16 implementation. Notably, even the largest TinyLIC-Base model, which provides the strongest R--D performance, remains highly practical at high resolutions. On an RTX 5090, TinyLIC-Base achieves over 150 FPS for both encoding and decoding at 1080p, demonstrating sufficient throughput for high-refresh-rate 120/144-FPS visual streams, such as interactive gaming and low-latency visual communication. At 4K, it still reaches 35.5 FPS for encoding and 49.0 FPS for decoding, maintaining real-time processing capability even for the most computationally demanding model variant. The smaller Nano and Small variants provide further throughput margins, making them suitable for higher-frame-rate or more resource-constrained deployment scenarios.

The practical significance is further reflected in comparison with conventional coding. TinyLIC-Base not only surpasses VTM-22 in R--D performance, but also achieves substantially higher runtime throughput, particularly for encoding and at commonly used resolutions such as 1080p. We note, however, that VTM-22 is reference software designed primarily for standard verification and algorithmic evaluation, rather than a highly optimized deployment implementation. Therefore, this comparison should not be interpreted as a comprehensive speed comparison against production-grade VVC codecs. Highly optimized and parallel VVC implementations, such as VVenC~\citep{vvenc} and VVdeC~\citep{vvdec}, as well as commercial hardware/software codecs, can achieve considerably higher throughput than
VTM-22. Nevertheless, the ability of TinyLIC to simultaneously provide real-time 4K processing, high-frame-rate 1080p throughput, and superior R--D performance demonstrates its strong potential for practical deployment.

\subsection{Discussion and Perspective}
\label{sec:tinylic_discussion}

The TinyLIC benchmark illustrates that practical learned image coding is
inherently a multi-objective design problem. Compression efficiency,
computational complexity, rate adaptability, numerical consistency, and
robustness are coupled through the underlying representation and entropy
model. Evaluating these properties under a shared architecture therefore
makes their interactions more explicit than comparisons among independently
developed codecs.

Three observations emerge from the benchmark. First, the
R--D--complexity trade-off can be adjusted over a relatively wide range
without changing the fundamental codec topology. The Nano, Small, and Base
configurations show that increasing model capacity consistently improves
coding efficiency, although the marginal gain decreases at higher
complexities. Complexity scalability should therefore be regarded as a
codec-design dimension rather than merely an implementation metric.

Second, different deployment requirements introduce different costs.
Variable-rate coding and selective quantization can be incorporated with
limited R--D degradation, whereas full integer inference imposes a stronger
numerical constraint. Progressive and loss-resilient coding further alter
the structure or dependency of the transmitted representation and therefore
introduce more visible compromises in nominal coding efficiency. The cost
of practical functionality thus depends strongly on where the corresponding
constraint is introduced into the coding pipeline.

Third, the experiments highlight the distinction between algorithmic
efficiency and practical usability. A codec with the strongest unconstrained
R--D performance is not necessarily the most suitable operating point once
decoding complexity, numerical reproducibility, or communication reliability
is considered. Practical comparison therefore extends beyond a conventional
R--D curve toward a broader rate--distortion--complexity--reliability
design space.

TinyLIC provides a common reference for studying this design space rather
than defining a single optimal solution. Different applications may
prioritize different properties, such as low decoding complexity,
bit-exact interoperability, flexible bitrate control, or resilience to
unreliable transmission. The role of the benchmark is to make the
corresponding trade-offs measurable under a controlled codec architecture.

\section{Concluding Remarks and Future Perspectives}
\label{sec:conclusion}

\subsection{Concluding Remarks}

End-to-end LIC has progressed from proof-of-concept neural
compression toward increasingly complete codec systems. Improvements in
learned transforms, quantization, entropy modeling, and end-to-end
optimization have substantially advanced R--D performance, but practical
adoption requires a broader set of properties. Rate adaptability,
deterministic decoding, robustness, computational efficiency, and
interoperability are becoming integral components of codec design rather
than secondary implementation considerations.

This survey has reviewed LIC from this deployment-oriented perspective,
treating the core transform--quantization--entropy--optimization framework
as the algorithmic foundation and focusing on the practical capabilities
required beyond nominal R--D optimization. The TinyLIC benchmark further
provides a common setting in which representative deployment constraints can
be evaluated under controlled architectural and computational conditions.

The standardization of JPEG AI and IEEE 1857.11 further demonstrates that
learned image coding is entering a stage of practical validation
\citep{JPEG-AI,alshina2024jpeg,IEEE1857}. Bitstream conformance, decoder
complexity, numerical reproducibility, and implementation interoperability
are becoming first-class codec requirements. These developments provide the
basis for the broader evolution of learned coding discussed next.

\subsection{Future Perspectives}

While the preceding sections focus primarily on making current LIC
architectures deployable, several emerging developments may further change
what visual coding is expected to represent and support. We highlight four
directions that are likely to influence the next stage of learned coding.

\subsubsection{Toward Coding for Intelligence}
\label{sec:future_intelligence}

The conventional objective of image coding is to reconstruct a signal for
human viewing. Future visual communication systems, however, will
increasingly deliver the same visual information to both humans and
machines. This motivates \emph{coding for intelligence}, in which the coded
representation should support not only high-quality reconstruction but also
recognition, analysis, and decision-making
\citep{zhang2024recent,yang2024video}.

Existing studies have explored task-oriented compression and unified
human--machine representations
\citep{zhang2024all,zhang2025perception}. Nevertheless, many current methods
remain dependent on particular downstream networks or tasks. Jointly
training the codec with a detector or classifier can provide strong task
performance, but it reduces generality and makes the representation
sensitive to changes in downstream models.

A more scalable objective is therefore to construct a common compressed
representation that can serve multiple consumers. Instead of maintaining
different bitstreams for human reconstruction, object recognition, and
semantic understanding, future codecs may expose a shared representation
that can be selectively decoded or adapted according to the requested task.
Such a design would extend the scalability discussed in Sec.~3 from
different \emph{rates} toward different \emph{uses} of the same coded
information.

The key challenge is determining what information should be preserved when
the downstream task is not known at encoding time. Signal fidelity,
perceptual realism, and machine utility do not necessarily assign the same
importance to visual information. Developing representations and evaluation
protocols that preserve broadly useful information while maintaining
bitstream interoperability therefore remains an open problem. Recent
discussions connecting efficient representation with biological or
machine intelligence provide an interesting conceptual motivation
\citep{zheng2025unbearable,huang2024compression}, but translating this idea
into general-purpose coding systems requires substantially stronger
empirical foundations.

\subsubsection{Beyond Latent-Grid Representations}
\label{sec:future_representation}

Most current LIC methods represent an image using spatial latent tensors
decoded by a shared neural network. Emerging neural representations broaden
this design space. Implicit neural representations (INRs), for example,
encode a signal through the parameters or latent variables of a
coordinate-conditioned function
\citep{sitzmann2020implicit}. This representation can provide continuous
spatial querying and naturally support resolution-flexible reconstruction.

Recent work has explored INRs and related overfitted neural representations
for image, video, and scientific-data compression
\citep{strumpler2022implicit,kim2024c3,shi2024hiner,shi2025quantizing,
ladune2026cool}. In parallel, explicit neural scene representations such as
Gaussian primitives have become effective representations for view
synthesis and immersive media
\citep{kerbl20233d}. These developments suggest that future visual codecs
may no longer be restricted to transmitting a conventional latent feature
map, but may instead encode functions, primitives, or other scene-level
representations.

Such representations also introduce different deployment trade-offs.
Instance-specific optimization can reduce the amount of transmitted data but
increase encoding time; transmitting model or representation parameters
requires new entropy models and numerical formats; and random access,
hardware acceleration, and deterministic decoding may differ substantially
from those of conventional VAE-based codecs. Their practical value should
therefore be assessed using the same deployment criteria emphasized
throughout this survey rather than compression performance alone.

\subsubsection{Foundation-Model-Assisted Compression}
\label{sec:future_foundation}

Foundation models provide another possible shift in the coding paradigm.
Large vision and multimodal models contain powerful priors for visual
structure and semantics, while continuing progress in model compression and
edge inference is making their deployment increasingly feasible
\citep{zheng2025review}. Instead of explicitly transmitting every visual
detail, a codec may transmit compact structural, semantic, or latent
information and rely on a shared generative prior to reconstruct plausible
content.

Recent studies already demonstrate the potential of this direction.
Large models can provide strong probability models for lossless compression
across different data modalities
\citep{li2025lossless}, while generative image codecs can reconstruct
visually realistic images at very low rates
\citep{careil2023towards,keultra}. Such approaches may be particularly
attractive in extremely bandwidth-constrained scenarios, where preserving
all pixel-level information is infeasible.

The resulting coding objective, however, differs fundamentally from
conventional fidelity-oriented compression. A generative reconstruction can
be perceptually convincing while differing semantically from the original
signal. Hallucination, semantic consistency, controllability, and
uncertainty therefore become codec-level concerns rather than purely
generative-model concerns. In applications such as archival storage,
medical imaging, remote sensing, or machine decision-making, plausible
reconstruction cannot automatically be treated as correct reconstruction.

Foundation-model-assisted compression will consequently require evaluation
beyond conventional R--D and perceptual quality. Future protocols may need
to characterize semantic fidelity and reconstruction reliability together
with bitrate and computational cost. Efficient deployment is equally
important: a reduction in transmitted bits is of limited practical benefit
if reconstruction requires an excessively large model or substantial
inference latency.

\subsubsection{From Learned Image Coding to Deployable Video Coding}
\label{sec:future_video}

Many principles reviewed in this article naturally extend from image to
video coding, but their system implications become substantially more
difficult. Video introduces temporal prediction, reference management,
buffer constraints, random access, and continuous real-time processing.
Consequently, numerical inconsistency or reconstruction error is no longer
confined to a single coding instance but may propagate through subsequent
frames.

Recent work has demonstrated increasingly practical neural video codecs,
including real-time mobile implementations
\citep{van2024mobilenvc} and initial studies of cross-platform deterministic
execution
\citep{conceiccao2025cross}. Nevertheless, practical learned video coding
must simultaneously satisfy substantially tighter constraints on latency,
memory bandwidth, energy consumption, and temporal consistency than
still-image coding.

Rate adaptation also becomes a sequential system problem. Network
conditions and frame-level coding decisions interact over time, connecting
video compression with adaptive streaming and communication control
\citep{chen2024toward,xu2023karma,li2023mamba}. In addition, decoder
complexity and reference dependencies limit the degree to which
high-capacity image-coding architectures can be transferred directly to
real-time video.

For these reasons, still-image LIC is likely to remain an important
testbed for deployment techniques before they become mature in learned
video coding. Advances in efficient neural accelerators, deterministic
low-precision computation, lightweight temporal modeling, error control,
and standardization will be central to closing this gap
\citep{gao2025implementations}.

\subsection{Closing Perspective}

Learned image coding is moving from optimizing a neural compression model
toward designing a complete visual communication system. In the near term,
the most important challenge is to make compression gains reproducible,
controllable, efficient, and interoperable under real deployment
constraints. In the longer term, the representation itself may evolve as
coding becomes increasingly integrated with machine perception, neural
scene representations, and generative foundation models.

These two directions should not be viewed independently. New
representations and more powerful priors will only become useful coding
technologies if their computational cost, numerical behavior, robustness,
and interoperability can also be controlled. Conversely, the deployment
mechanisms developed for current LIC provide an important foundation for
future AI-native visual coding systems. The next generation of learned
codecs will therefore be shaped not by R--D performance alone, but by the
joint optimization of \emph{compression efficiency, practical
deployability, and intelligence compatibility}.

\section*{Declarations}

\subsection*{Availability of data and materials}
The datasets used in this study are publicly available from their respective
sources. The implementation, pretrained models, and benchmark materials of
TinyLIC are available at
\url{https://njuvision.github.io/TinyLIC}.

\subsection*{Competing interests}
The authors declare that they have no competing interests.

\subsection*{Funding}
This work was supported in part by the National Natural Science Foundation of China under Grant No. 62431011.

\subsection*{Authors' contributions}
Ming Lu contributed to manuscript drafting. Junqi Shi
contributed to experimental studies and the open benchmark. Zhan Ma contributed
to manuscript polishing and resource support. All authors reviewed and approved the
manuscript.

\subsection*{Acknowledgements}
Not applicable.

\bibliography{e2e_cleaned}% common bib file

\end{document}